\documentclass{amsart}

\pdfoutput=1

\usepackage{amsmath}
\usepackage{thm-restate}
\usepackage{graphicx}

\newtheorem{theorem}{Theorem}[section]
\newtheorem{lemma}[theorem]{Lemma}

\theoremstyle{definition}
\newtheorem{Prob}[theorem]{Problem}
\theoremstyle{remark}
\newtheorem{remark}[theorem]{Remark}

\newcommand{\bra}{\langle}
\newcommand{\ket}{\rangle}
\newcommand{\Ex}{\mathrm{E}}

\usepackage{amsaddr}

\title[]{A Faster Subquadratic Algorithm\\ For Finding Out\-lier Cor\-re\-la\-tions}
\author[]{Matti Karppa, Petteri Kaski, and Jukka Kohonen}
\address{Aalto University, Department of Computer Science}


\begin{document}

\begin{abstract}
We study the problem of detecting {\em outlier pairs} 
of strongly correlated variables among a collection of $n$ variables 
with otherwise weak pairwise correlations.
After normalization, this task amounts to the geometric task where 
we are given as input a set of $n$ vectors with unit Euclidean norm
and dimension $d$, and for some constants $0<\tau<\rho<1$, we are asked to find all the outlier pairs of 
vectors whose inner product is at least $\rho$ in absolute value, subject 
to the promise that all but at most $q$ pairs of vectors have inner product 
at most $\tau$ in absolute value. 

Improving on an algorithm of G.~Valiant [FOCS~2012; J.\,ACM~2015],
we present a randomized algorithm that for Boolean inputs 
($\{-1,1\}$-valued data normalized to unit Euclidean length)
runs in time 
\[
\tilde O\bigl(n^{\max\,\{1-\gamma+M(\Delta\gamma,\gamma),\,M(1-\gamma,2\Delta\gamma)\}}+qdn^{2\gamma}\bigr)\,, 
\]
where $0<\gamma<1$ is a constant tradeoff parameter and $M(\mu,\nu)$ is the  
exponent to multiply an $\lfloor n^\mu\rfloor\times\lfloor n^\nu\rfloor$ matrix with an $\lfloor n^\nu\rfloor\times \lfloor n^\mu\rfloor$ 
matrix and \mbox{$\Delta=1/(1-\log_\tau\rho)$}.
As corollaries we obtain randomized algorithms that run in time
\[
\tilde O\bigl(n^{\frac{2\omega}{3-\log_\tau\rho}}+qdn^{\frac{2(1-\log_\tau\rho)}{3-\log_\tau\rho}}\bigr)
\]
and in time 
\[
\tilde O\bigl(n^{\frac{4}{2+\alpha(1-\log_\tau\rho)}}+qdn^{\frac{2\alpha(1-\log_\tau\rho)}{2+\alpha(1-\log_\tau\rho)}}\bigr)\,,
\]
\baselineskip=9.25pt
where $2\leq\omega<2.38$ is the exponent for square matrix multiplication and 
\mbox{$0.3<\alpha\leq 1$} is the exponent for\, rectangular matrix multiplication. 
The notation {$\tilde O(\cdot)$} hides 
polylogarithmic factors in $n$ and $d$ whose degree may depend on 
$\rho$ and $\tau$.
We present further corollaries for the light bulb problem and for learning 
sparse Boolean functions. 
\end{abstract}

\maketitle

\section{Introduction}

\label{sect:introduction}

\subsection{Scalable Correlation Analysis}

The Pearson product-moment correlation coefficient~%
\cite{Galton1888,Pearson1920,Stigler1989},
or briefly, {\em correlation}, is one of the most fundamental statistical 
quantities used to measure the strength of interaction between two 
sequences of $d$ numerical observations. Assuming we have $n$ such sequences, 
there are $\Theta(n^2)$ pairs of sequences, which immediately puts forth
the algorithmic question of scalability of correlation-based 
analyses as our number of observables $n$ increases.

If we seek to explicitly compute all $\Theta(n^2)$ pairwise
correlations, the size of our desired result forces us 
to quadratic $\Omega(n^2)$ scaling as a function of $n$. 
Thus, assuming we seek subquadratic%
\footnote{For brevity, we will use the expression
``subquadratic in $n$'' to mean $O(n^{2-\epsilon})$ for
some constant $\epsilon>0$ independent of $n$.}
scaling in $n$, we must refine our objective towards aggregate analyses 
that only implicitly consider all the pairwise correlations. One natural goal 
in this setting is to identify interesting pairs of observables, 
subject to the assumption that there are few such pairs. 
In the context of correlation, one possible signal for an interesting 
interaction between a pair of observables is an abnormally large, 
or {\em outlier}, (absolute) correlation, as measured against a 
{\em background} parameter that bounds from above the absolute values 
of correlations between most pairs of observables.

Assuming our $n$ observables are each normalized to zero mean and 
unit standard deviation, the computational task of finding abnormally large 
correlations reduces to the following geometric setup:

\begin{Prob}[\textsc{Out\-lier Cor\-re\-la\-tions}]
\ Given as input a set of $n$ vectors with unit Euclidean norm
and dimension $d$, find all the {\em outlier pairs} 
of vectors whose inner product is at least $\rho$ in absolute value, 
subject to the {\em promise} that all but at most $q$ pairs of 
vectors have inner product at most $\tau$ in absolute value, for 
some $0<\tau<\rho<1$.
(The promise implies in particular that there are at most $q$ outlier pairs.
Our interest is on inputs where $q$ is subquadratic in $n$.)
\end{Prob}

\begin{remark}
A useful variant of the problem is that the input vectors are each
assigned one of two possible colors, and we want to find all the outlier
pairs with distinct colors, subject to the same promise as above. 
Let us call this variant \textsc{Bichromatic Out\-lier Cor\-re\-la\-tions}.
In what follows we can work with the bichromatic variant since 
any instance of \textsc{Outlier Correlations} can be made 
bichromatic by a routine reduction with a multiplicative polylogarithmic
cost in $n$.%
\footnote{Suppose the data matrix 
for \textsc{Out\-lier Cor\-re\-la\-tions} is $D\in\{-1,1\}^{d\times n}$ and 
we are interested in discovering the pairs of distinct columns 
in $D$ with (absolute) inner product at least $\rho d$. 
Construct $\lceil\log_2 n\rceil$ pairs of matrices $A_k,B_k$ 
for $k=0,1,\ldots,\lceil\log_2 n\rceil-1$, where the matrix 
$A_k$ (respectively, $B_k$) contains the columns of $D$ whose index, 
viewed as a $\lceil\log_2 n\rceil$-bit string, has a 0 (respectively, 1) 
in bit position $k$. Take the union of the solutions of the instances 
$A_k,B_k$ to obtain the solution for $D$. All distinct columns will
be discovered because each pair of distinct columns differs in at least
one bit position $k$.}{}
The bichromatic variant in particular captures a batch-query database setting, 
where the vectors with one color constitute a database, the vectors with the
other color constitute a batch of queries to the database, and the 
outlier pairs with distinct colors capture correlated matches in
the database to the queries.
\end{remark}

The question of scalability as $n$ increases is now parameterized by 
the dimension $d$ of the vectors and the thresholds $\rho,\tau$.
Furthermore, we can impose restrictions on the structure of the vectors 
themselves to study restricted variants of the problem.
From a discrete computational standpoint, a natural restriction 
is to assume the input vectors take values in $\{-1,1\}$.%
\footnote{%
An arbitrary input on the unit Euclidean sphere can be 
quantized to a $\{-1,1\}$-valued input if one is willing to accept 
randomization and an expected loss 
$\lambda\mapsto 1-\frac{2}{\pi}\arccos\lambda$ in the
parameters $\lambda\in\{\tau,\rho\}$ caused by rounding 
the arbitrary input into a Boolean input with random hyperplanes. 
See~e.g.~\cite[Algorithm~4]{Valiant2015} or \cite{Charikar2002}.}
(Indeed, two vectors in $x,y\in\{-1,1\}^d$ have
inner product $\bra x,y\ket=\sum_{i=1}^d x_iy_i=\lambda d$ if 
and only if the corresponding normalized vectors have inner product 
$\bra x/\sqrt{d},y/\sqrt{d}\ket=\lambda$ with $-1\leq\lambda\leq 1$.)
Let us call $\{-1,1\}$-valued inputs {\em Boolean} inputs.

\begin{remark}
For Boolean vectors $x,y\in\{-1,1\}^d$, the inner product and 
the {\em Hamming distance} $D_H(x,y)=|\{i\in\{1,2,\ldots,d\}:x_i\neq y_i\}|$ 
are affinely equivalent via the identity 
\begin{equation}
\label{eq:hamming-ip}
\bra x,y\ket + 2D_H(x,y)=d\,. 
\end{equation}
Thus, for Boolean inputs the \textsc{Out\-lier Cor\-re\-la\-tions} problem is 
a variant of {\em near neighbor search} under the Hamming metric%
---the larger the inner product, the smaller the Hamming distance,
and vice versa. (Cf.~\cite{Indyk2004,Minsky1969,Zezula2006}.)
\end{remark}

\subsection{Earlier Lower Bounds and Upper Bounds}
\label{sect:lower-upper}

Already for Boolean inputs it is known that a quest for sub\-quadratic 
scaling in $n$ is faced with fine-grained complexity-theoretic barriers. 
Indeed, unless the Strong Exponential Time Hypothesis~\cite{Impagliazzo2001} 
is false, any algorithm that for a bichromatic input detects the existence of 
an outlier pair with $n^{2-\epsilon}$ scaling in $n$ 
(and simultaneous $2^{o(d)}$ scaling in the dimension $d$) 
must have the property that $\epsilon$ depends on $\rho$ and $\tau$. 
More precisely, it follows by a simple local transformation 
(see~\S\ref{sect:orthogonal-vectors}) from a result of 
Williams~\cite{Williams2005} that unless SETH is false, $\epsilon$ cannot 
be bounded from below by a positive constant when $|\rho-\tau|\rightarrow 0$.

On the positive side, algorithms for the 
$(1+\epsilon)$-{\em approximate nearest neighbor} problem can be used 
to identify outlier pairs.%
\footnote{%
By \eqref{eq:hamming-ip}, two Boolean vectors with 
inner product $\lambda d$ have Hamming distance $(1-\lambda)d/2$. 
Thus, assuming there are no pairs whose inner product is 
strictly between $\tau$ and $\rho$, the vectors that
belong to outlier pairs can be discovered by selecting
a small enough $\epsilon$ so that an $(1+\epsilon)$-approximate 
nearest neighbor of a vector has to be a completion of such a
vector to an outlier pair if such a completion exists. For example,
by \eqref{eq:hamming-ip} we can take $1+\epsilon<(1-\tau)/(1-\rho)$ for 
the Hamming metric and $(1+\epsilon)^2<(1-\tau)/(1-\rho)$ for 
the Euclidean metric.}
\textit{Locality-sensitive hashing} (LSH), introduced by Indyk and
Motwani~\cite{Indyk1998}, is a well-known framework for finding
$(1+\epsilon)$-approximate nearest neighbors. 
In their seminal paper, 
Indyk and Motwani describe an algorithm that, for $\epsilon > 0$ and 
the Hamming metric, uses $\tilde O(n^{1+1/(1+\epsilon)}+dn)$ space 
for supporting approximate nearest-neighbor queries to a data set of 
$n$ vectors of dimension $d$, and requires $\tilde O(dn^{1/(1+\epsilon)})$ 
query time.%
\footnote{The notation $\tilde O(\cdot)$ suppresses factors 
polylogarithmic in $n$ and $d$ whose degree may depend on $\rho$ and $\tau$.}
Lower bounds are known for the hash function
families~\cite{Motwani2007,ODonnell2014} which guarantee that
algorithms based on data-independent LSH, such as Andoni and 
Indyk~\cite{Andoni2006} with query time $dn^{1/(1+\epsilon)^2+o(1)}$
for the Euclidean metric, are essentially optimal.
Andoni, Indyk, Nguyen and Razenshteyn~\cite{Andoni2014} 
present a framework that circumvents these lower bounds
by introducing data dependence into the hash families, 
giving the improved query exponent 
$7/(8(1+\epsilon)^2)+O(1/(1+\epsilon)^3)+o(1)$ 
for the query time in the Euclidean metric.
Andoni and Razenshteyn~\cite{Andoni2015} optimize this
exponent to $1/(2(1+\epsilon)^2-1)+o(1)$ and
Andoni {\em et al.}~\cite{AILRS2015} 
present LSH optimized for angular distance.

Turning from approximate to exact solutions,
Alman and Williams~\cite{Alman2015}
show that the {\em batch} Hamming nearest neighbor problem with $n$ 
simultaneous queries to a database of $n$ vectors can be solved in randomized
$n^{2-1/O(c(n)\log c(n))}$ time if $d=c(n)\log n$ for 
$c : \mathbb{N}\rightarrow\mathbb{N}$. 
This scaling is subquadratic 
in $n$ if the dimension $d$ is bounded by $d=O(\log n)$.

\subsection{The Curse of Weak Outliers and Valiant's Breakthrough}

For Boolean inputs with unrestricted dimension, the scaling obtained 
via approximate nearest neighbors is subquadratic in $n$ for all constant 
$\rho>\tau$. However, such scaling suffers from what could be called the
{\em curse of weak outliers}. Namely, if $\rho$ is small 
(that is, the outlier correlations themselves are weak), 
the scaling in $n$ is essentially quadratic $n^{2-c\rho}$ for 
a positive constant $c$, {\em even if the background $\tau$ decays to zero}.
Ideally, there should be a way to avoid such direct dependence 
on the value of $\rho$, as long as $\rho$ and $\tau$ do not converge.

In a breakthrough result, G.~Valiant \cite{Valiant2015} showed 
that subquadratic scaling {\em is} possible,
essentially independently of $\rho$, as long as $\log_\tau\rho=(\log\rho)/(\log\tau)$ 
remains bounded from above by a small positive constant.
Here the logarithmic ratio $\log_\tau\rho$ quantifies the gap between 
the background and the outliers; since
$0<\tau<\rho<1$, both logarithms are negative, and the 
ratio provides an essentially inverted quantification of the gap so that 
the gap is wide when the ratio is close to~0 and narrow when the ratio 
is close to~1 (cf.~Figure~\ref{fig:runtimes_current}).

Let $2\leq\omega<2.3728639$ be the exponent%
\footnote{%
We adopt the convention that all running time bounds that contain 
an exponent $f(\omega)$ of $n$ that depends on $\omega$ (or any other 
limiting exponent for matrix multiplication, such as $\alpha$) are tacitly stated as 
$f(\omega)+\epsilon$ for an arbitrarily small constant $\epsilon>0$.}
of square matrix multiplication~\cite{LeGall2014}. 
We say that a parameterized event $E_n$ happens 
{\em with high probability (w.h.p.) in the parameter $n$} 
if $\Pr(E_n)\geq 1-f(n)$ for a function $f(n)$ with 
$\lim_{n\rightarrow\infty} n^k f(n)=0$ for all positive $k$.

Valiant's algorithm runs in two phases. The first one is the
\emph{approximate detection}\footnote{We may stop the algorithm after the approximate detection phase
if we only want to decide whether the input contains at least
one pair with absolute inner product at least $\rho d$, with the
following approximation guarantees: 
(i) if all the inner products in the input have absolute 
value at most $\tau d$, the algorithm outputs false w.h.p.; and 
(ii) if the input contains at least one inner product with 
absolute value at least $\rho d$, the algorithm outputs true w.h.p.}{}
phase where the input vectors are divided into
blocks and it is decided which pairs of blocks contain one or more
vector pairs with
absolute inner product in the excess of $\rho d$, if any. The second
phase is the \emph{listing} phase where a brute-force search is
performed on the pairs corresponding to the indicated blocks.

\begin{theorem}[Valiant~\cite{Valiant2015}]
\label{thm:valiant-main}
For all constants $0<\tau<\rho<1$, 
the \textsc{Out\-lier Cor\-re\-la\-tions} problem for Boolean inputs
admits a randomized algorithm that runs in 
time $\tilde O(n^{\frac{5-\omega}{4-\omega}+\omega\log_\tau\rho})$
for approximate detection and subsequent $\tilde O(qdn^{1/(4-\omega)})$ time 
for exact listing of all the outliers, w.h.p.
\end{theorem}

\begin{remark}
For any constant $\rho>0$, as $\tau\rightarrow 0$ 
we observe that
the exponent for approximate detection in Theorem~\ref{thm:valiant-main} 
approaches $1.50\leq (5-\omega)/(4-\omega)<1.62$. 
Thus, Valiant's algorithm gives subquadratic scaling for all constant $\rho$, 
as long as $\tau$ is comparatively smaller.
\end{remark}

\begin{remark}
A tacit property of Valiant's algorithm is that the running time 
for approximate detection is, up to polylogarithmic factors, 
independent of the dimension $d$ of the input. 
In particular, since the input is Boolean, the (normalized) input vectors 
are decoherent
and hence it suffices to randomly sample the input to access
the correlations. By {\em decoherence} we mean here the property that 
if we view the squares of the entries of a (normalized) input vector as 
a probability distribution over the $d$ dimensions, 
the probability mass is roughly uniformly distributed across 
the dimensions. With (normalized) Boolean input, the distribution
is exactly uniform and thus eases concentration analyses.
\end{remark}

Valiant's result opens up a quest to understand the extent of 
subquadratic scaling available for \textsc{Out\-lier Cor\-re\-la\-tions}:

\smallskip
\noindent
{\em Assuming that the outliers are well-separated from the background
correlations, that is, that $\log_\tau\rho$ is small, 
how close to linear scaling in $n$ can we get?}

\smallskip \noindent 
Our intent in this paper is to present further progress in 
such a quest.

\subsection{Valiant's Algorithm in More Detail}
\label{sect:valiant-algorithm}

Since our intent is to improve on Valiant's algorithm, 
it will be convenient to review the key ideas in its design.

Let us assume the input of our instance of 
\textsc{Bi\-chro\-mat\-ic Out\-lier Cor\-re\-la\-tions} is given to us
as two $d\times n$ Boolean 
matrices, $A\in\{-1,1\}^{d\times n}$ and \mbox{$B\in\{-1,1\}^{d\times n}$}.
Our task is to find the outlier inner products 
(of absolute value at least $\rho d$) between columns of $A$ and $B$.
Valiant's algorithm proceeds in the following four phases; 
the first and second phases together constitute a compression phase 
which enables the use of fast matrix multiplication in the third 
approximate detection phase, which is followed by an exact listing phase.

{\em 1. Expansion by uniform random sampling of a tensor power.} 
It is possible to amplify
the inner products between columns of $A$ and $B$
by individually tensoring the matrices $A$ and $B$,
at the cost of increasing the dimension of the data from $d$ to $d^p$ for 
a positive integer $p$. Indeed, for any two $d$-dimensional real vectors,
it holds that 
$\bra x,y\ket^p=\bra x^{\otimes p},y^{\otimes p}\ket$, where $\otimes p$ 
indicates $p$-fold Kronecker product (tensoring) of the vector with itself.
Accordingly, taking $p$-fold Kronecker products in the vertical dimension 
only (which we indicate with the notation $\downarrow\!\!\!\otimes p$)
we obtain the $d^p\times n$ matrices $A^{\downarrow\otimes p}$ and 
$B^{\downarrow\otimes p}$. A key observation is that a uniform random 
sample of size $s$ (the value of $s$ will be fixed later) of the $d^p$ dimensions suffices 
because the input is decoherent (Boolean) and hence the sum of the 
sample is strongly concentrated, for example, via the \mbox{Hoeffding}
bounds~\cite{Hoeffding1963}.
Accordingly, we assume that the matrices $A^{\downarrow\otimes p}$ 
and $B^{\downarrow\otimes p}$ in fact have dimensions $s\times n$.

{\em 2. Signed aggregation.}
Since comparatively few of the $n^2$
inner products between the columns of $A^{\downarrow\otimes p}$ and 
$B^{\downarrow\otimes p}$ are outliers, after amplification it is possible 
to aggregate the $n$ columns in $A^{\downarrow\otimes p}$ 
(respectively, $B^{\downarrow\otimes p}$) by randomly 
partitioning the columns into $n/t$ blocks of $t$ columns, and taking, 
with an independent uniform random sign for each column, 
the sum of the columns in each block. This produces an
$s\times (n/t)$ matrix $\tilde A$ (respectively, $\tilde B$). 

{\em 3. Approximate detection.}
Because the outer dimension has now decreased from $n$ to $n/t$, with
careful selection of the parameters $s,t,p$ we can now afford to 
multiply the compressed matrices $\tilde A,\tilde B$ in subquadratic time
to obtain the $(n/t)\times(n/t)$ product $\tilde A^\top\tilde B$. 
Because of careful amplification by sampling and signed aggregation, 
the pairs of blocks containing at least one outlier pair are with 
moderate probability signalled by entries in $\tilde A^\top\tilde B$ 
that have absolute value above a threshold value.%
\footnote{%
The algorithm is iterated $\Theta(\log n)$ times to guarantee that
every outlier pair is signalled in at least one pair of blocks 
during at least one iteration with high probability.}{}

{\em 4. Exact listing.}
Finally, the outlier pairs can be computed with brute force by 
computing the $t^2$ pairwise inner products of columns of $A$ and $B$
within each signalled pair of blocks.

Theorem~\ref{thm:valiant-main} follows by careful selection of 
the parameters $s,t,p$ and the signalling threshold.

\subsection{Our Contribution}
\label{sect:contribution}

Our contribution in this paper amounts to the observation
that the compression phase (expansion followed by signed aggregation) 
in Valiant's algorithm can be replaced by a faster compression 
subroutine. In essence, we rely on fast matrix multiplication
as the algorithmic device to {\em simultaneously} expand and aggregate; 
this requires that we replace the uniform random sampling of dimensions 
in the expansion phase of Valiant's algorithm with 
Cartesian product sampling to enable us to entangle 
expansion and aggregation. 
Despite the considerable decrease in entropy compared with a uniform 
random sample, we show that Cartesian product sampling 
{\em on a tensor power} remains roughly as sharply concentrated as 
a uniform random sample, enabling us to use roughly the same sample 
size $s$ as Valiant's algorithm for comparable compression, thus 
resulting in faster execution because of speedup given by fast 
matrix multiplication.
The faster compression subroutine then enables a faster 
tradeoff that balances between compression and detection, as controlled 
by a tradeoff parameter $0<\gamma<1$.

Let us now state our main result for \textsc{Out\-lier Cor\-re\-la\-tions}.
For constants $\mu,\nu>0$, let $M(\mu,\nu)$ be the infimum of
the values $\sigma>0$ such that there exists an algorithm that multiplies
an $\lfloor n^\mu\rfloor\times \lfloor n^\nu\rfloor$ integer matrix with 
an $\lfloor n^\nu\rfloor\times \lfloor n^\mu\rfloor$ integer matrix 
in $O(n^\sigma)$ arithmetic operations. For example, $\omega=M(1,1)$.

\begin{restatable}[Main]{theorem}{mainthm}
\label{thm:main}
For all constants $0<\gamma<1$ and $\Delta\geq 1$,
the \textsc{Out\-lier Cor\-re\-la\-tions} problem for Boolean inputs
admits a randomized algorithm that runs in 
time 
\begin{equation}
\label{eq:main-bound}
\tilde O\bigl(n^{\max\,\{1-\gamma+M(\Delta\gamma,\gamma),\,M(1-\gamma,2\Delta\gamma)\}}\tau^{-4}\bigr)
\end{equation}
for approximate detection and subsequent 
$
\tilde O\bigl(qdn^{2\gamma}\bigr)
$
time for exact listing of all the outliers, 
w.h.p.
The running time bounds hold uniformly for 
all $n^{-\Theta(1)}\leq\tau<\rho<1$ with 
$\log_\tau\rho\leq 1-\Delta^{-1}$.
\end{restatable}

For specific choices of $\gamma$ and $\Delta$ we obtain
the following corollaries.

\begin{restatable}{cor}{coromega}
\label{cor:omega}
For all constants $0<\tau<\rho<1$, 
the \textsc{Outlier Correlations} problem for Boolean inputs
admits a randomized algorithm that runs in time 
$
\tilde O\left(n^{\frac{2\omega}{3-\log_\tau\rho}}\right)
$
for approximate detection and subsequent 
$
\tilde O\left(qdn^{\frac{2(1-\log_\tau\rho)}{3-\log_\tau\rho}}\right)
$
time for exact listing of all the outliers, w.h.p.
\end{restatable}

Let $0.30298<\alpha\leq 1$ be the exponent for rectangular matrix 
multiplication~\cite{LeGall2012}. That is, $\alpha$ is the supremum
of all values $\sigma\leq 1$ with $M(1,\sigma)=2$.

\begin{restatable}{cor}{coralpha}
\label{cor:alpha}
For all constants $0<\tau<\rho<1$, 
the \textsc{Outlier Correlations} problem for Boolean inputs
admits a randomized algorithm that runs in time 
$
\tilde O\left(n^{\frac{4}{2+\alpha(1-\log_\tau\rho)}}\right)
$
for approximate detection and subsequent 
$
\tilde O\left(qdn^{\frac{2\alpha(1-\log_\tau\rho)}{2+\alpha(1-\log_\tau\rho)}}\right)
$
time for exact listing of all the outliers, w.h.p.
\end{restatable}

\begin{figure*}[t]
  \begin{center}
    \includegraphics[width=0.46\textwidth,trim={5mm 0mm 14mm 5mm},clip]{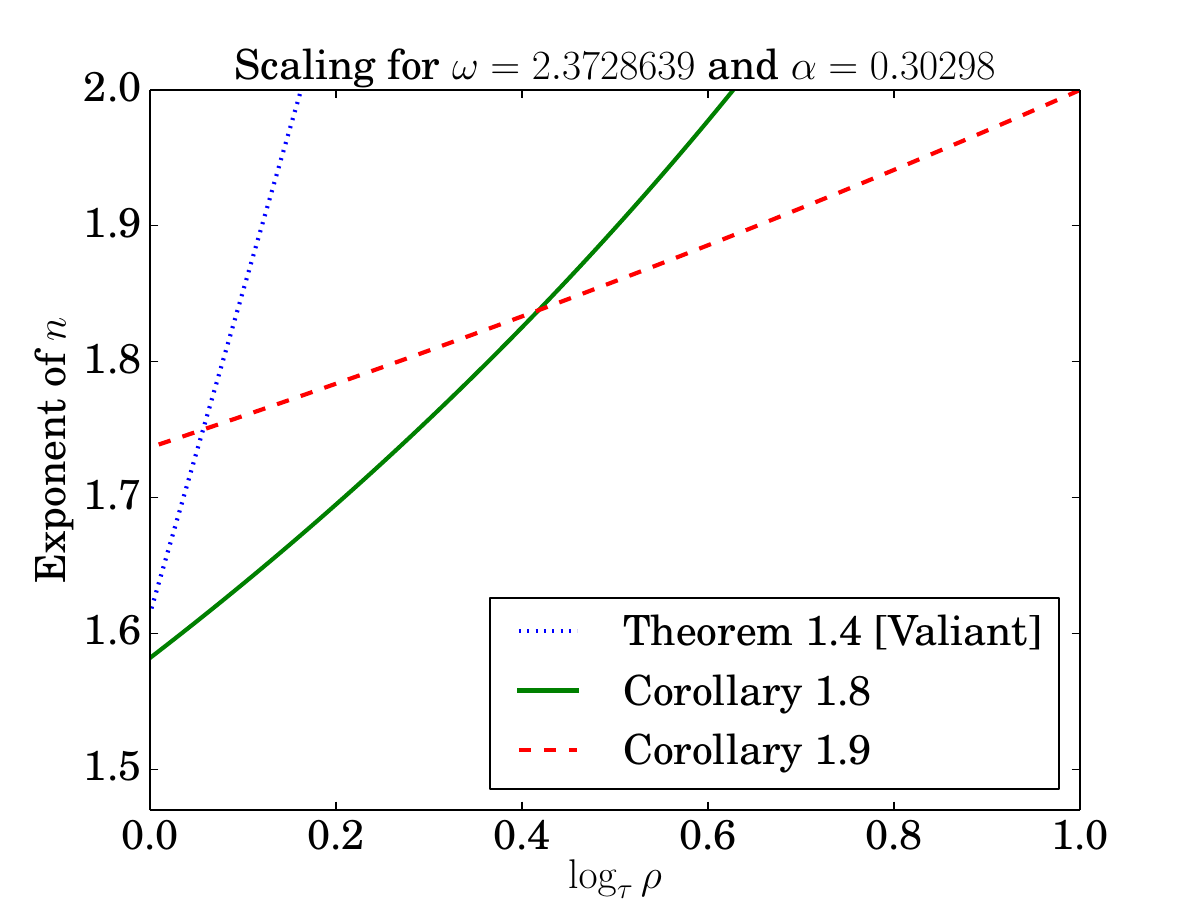}\hspace*{10mm}
    \includegraphics[width=0.46\textwidth,trim={5mm 0mm 14mm 5mm},clip]{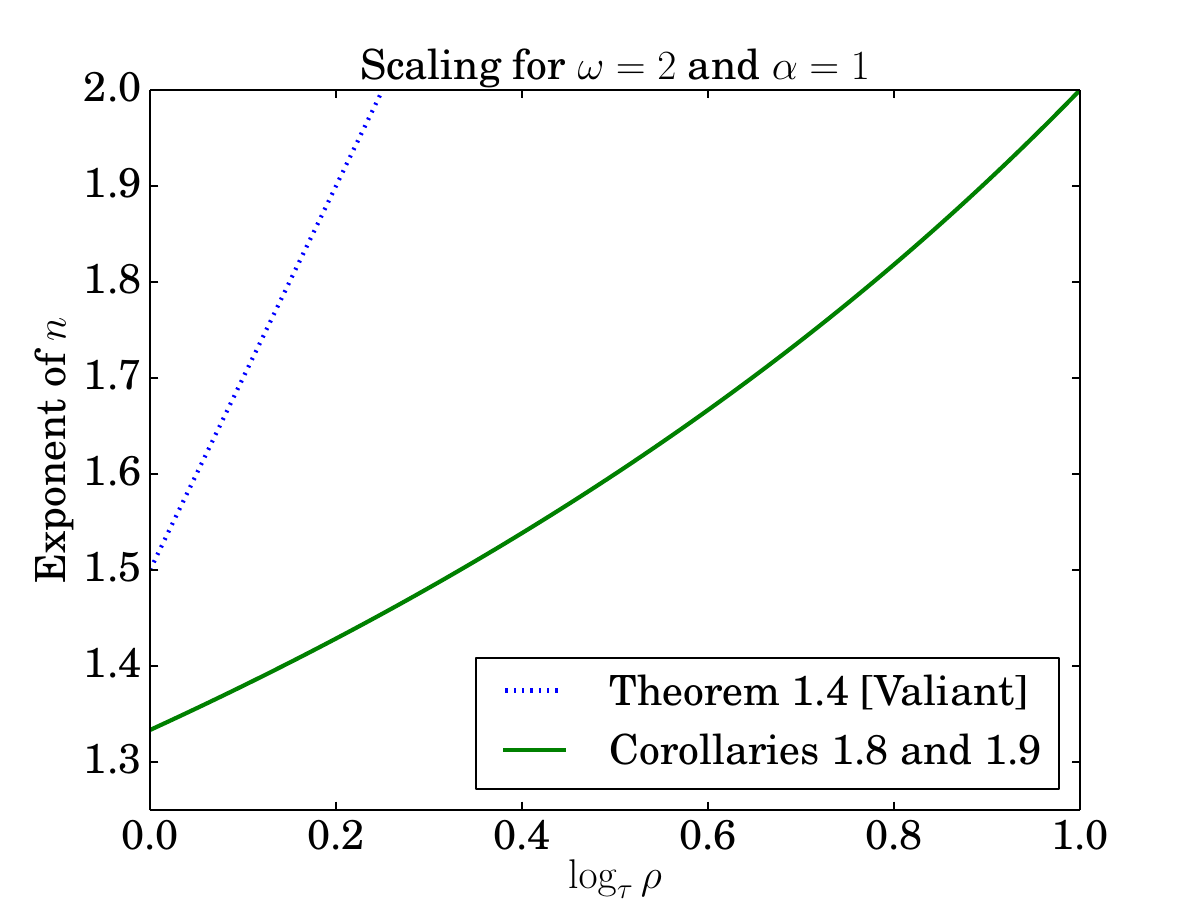}
    \caption{Illustration of subquadratic scaling in $n$ for approximate detection as a function of the parameter $\log_\tau\rho$. The matrix multiplication exponents $\omega$ and $\alpha$ are fixed to values $\omega=2.3728639$ and $\alpha=0.30298$ (left), and to values $\omega=2$ and $\alpha=1$ (right).}
    \label{fig:runtimes_current}
  \end{center}
\end{figure*}

Figure~\ref{fig:runtimes_current} displays the subquadratic scaling in $n$
for approximate detection obtained from Corollaries~\ref{cor:omega} 
and~\ref{cor:alpha}, contrasted with Theorem~\ref{thm:valiant-main} of~\cite{Valiant2015}.

\begin{remark}
\label{rem:weak}
Corollary~\ref{cor:alpha} implies that asymptotically it 
is possible to list extremely weak outliers with, say, 
$\rho=2^{-100}$ and $\tau=2^{-101}$, in time $\tilde O(n^{1.998}+qdn^{0.003})$.
\end{remark}

\begin{remark}
In the setting of well-separated outliers with any constant $\rho$
and \hbox{$\tau\rightarrow 0$}, the exponent for approximate detection 
in Corollary~\ref{cor:omega} approaches $2\omega/3$, improving 
Valiant's exponent $(5-\omega)/(4-\omega)$ across the range $2\leq\omega<2.38$.
\end{remark}

Although our results show that exponents as low as $2\omega/3=4/3$ may
be feasible for finding outlier correlations (if we assume $\omega=2$
and let $\tau\rightarrow 0$), the present framework unfortunately
appears not to be powerful enough to lower the exponent below
$4/3$. It remains open whether the exponent can be lowered all the way
to $1$ (to linear scaling in $n$), and whether techniques other than
fast matrix multiplication can be used to attain subquadratic scaling
without the curse of weak outliers.

\section{Related Work and Applications}

\label{sect:related-work}

The \textsc{Out\-lier Cor\-re\-la\-tions} problem 
has many applications, since the inner product of two vectors can be
used to measure the similarity of objects. For instance,~\cite{%
Achlioptas2011,%
Bayardo2007,%
Charikar2002,%
Chaudhuri2006,%
Cohen2001,%
Curtin2013,%
Das2007,%
Koenigstein2012,%
Lee2010,%
Low2012,%
Neyshabur2015,%
Ram2012,%
Shrivastava2014,%
Shrivastava2015,%
Silva2010,%
Teflioudi2015,%
Wang2011,%
Wang2012,%
Wang2013,%
Xia2004,%
Xiao2011,%
Zadeh2013%
}.
Here we will be content with discussing a narrow set of applications
of our results and related work. The proofs of all corollaries appear
in~\S\ref{sect:corollaries}. 

\subsection{The Light Bulb Problem}

Theorem~\ref{thm:main} gives as an almost immediate corollary 
a faster subquadratic algorithm (cf.~\cite[Corollary~2.2]{Valiant2015})
for solving the {\em light bulb problem} \cite{Valiant1988}, 
which asks us to discover a hidden correlated pair of light bulbs among
$n$ light bulbs blinking on and off independently and uniformly at random. 
Let us first state the problem in more precise terms and then give 
our improvement.

\begin{Prob}[\textsc{Light Bulb}]
Suppose we are given as input a set of $n$ vectors in $\{-1,1\}^d$
consisting of 
(i) a {\em planted pair} with inner product 
at least $\rho d$ in absolute value for $0<\rho<1$, and 
(ii) $n-2$ independent uniform random vectors in $\{-1,1\}^d$.
Our task is to find the planted pair among the $n$ vectors.%
\footnote{It follows from the Hoeffding 
bound~\eqref{eq:hoeffding} and the union bound that there is a 
constant $c>0$ such that for $d\geq c\rho^{-2}\log n$
with probability $1-o(1)$ the planted pair is the unique pair 
of vectors in the input with inner product at least $\rho d$ in absolute value.}
\end{Prob}
\begin{restatable}{cor}{corlightbulb}
\label{cor:light-bulb}
For all constants $0<\epsilon<\omega/3$, the \textsc{Light Bulb} problem
admits a randomized algorithm that for all 
$d\geq 5\rho^{-\frac{4\omega}{9\epsilon}-\frac{2}{3}}\log n $
runs in time 
\begin{equation}
\label{eq:light-bulb-bound}
\tilde O\left(n^{\frac{2\omega}{3}+\epsilon}\rho^{-\frac{8\omega}{9\epsilon}-\frac{4}{3}}\right)
\end{equation}
and finds the planted pair of vectors, with probability $1-o(1)$.
The running time bound holds uniformly for all $n^{-\Theta(1)}<\rho<1$.
\end{restatable}

Prior to Valiant's algorithm, the first subquadratic algorithm 
for the light bulb problem was the randomized 
$\tilde O(n^{1+(\log\frac{1+\rho}{2})/(\log\frac{1+\sigma}{2})})$ 
time algorithm of Paturi, Rajasekaran, and Reif~\cite{Paturi1989},
where $\sigma$ measures the absolute value of the inner product 
between the pair of vectors with the second largest inner product. 
Besides the algorithms based on locality-sensitive hashing within the
context of approximate nearest neighbors (see~\S\ref{sect:lower-upper}) that
can also be used to solve light bulb problem, the \textit{bucketing codes} 
approach of 
Dubiner~\cite{Dubiner2010}
yields a randomized 
$\tilde O(n^{2/(\rho+1)})$ time algorithm for the light bulb problem,
which was the fastest algorithm prior to Valiant's algorithm.
May and Ozerov~\cite{May2015}
present a recursive approach
for solving the light bulb problem; with weak outliers also their 
algorithm converges to quadratic running time.

\subsection{Learning Boolean Functions}

The light bulb problem generalizes to the task of
{\em learning a parity function in the presence of noise}.

\begin{Prob}[\textsc{Parity with Noise}]
Let the {\em support} $S\subseteq\{1,2,\ldots,n\}$ of the parity
function have size $k=|S|$ and let the {\em noise rate} be $0<\eta<1$. 
Our task is to find $S$, 
given access to independent {\em examples} of the form $(x,y)$, where 
(i) the {\em input} $x\in\{-1,1\}^n$ is chosen uniformly at random, and 
(ii) the {\em label} $y=z\cdot\prod_{j\in S}x_j$ is defined
by independently choosing $z\in\{-1,1\}$ with $\Pr(z=-1)=\eta$.
\end{Prob}%
The general case of unrestricted $k$ 
is studied by 
Blum, Kalai, and Wasserman~\cite{Blum2003} and
Lyubashevsky~\cite{Lyubashevsky2005}. 
Studies of the {\em sparse} case with $k=O(1)$
include works by Grigorescu, Reyzin, and Vempala
~\cite{Grigorescu2011} 
and Valiant~\cite{Valiant2015}. In the sparse case, 
we can use a split-and-list transformation presented by 
Valiant~\cite[p.~32]{Valiant2015} 
together with the algorithm underlying 
Corollary~\ref{cor:light-bulb} to essentially match an algorithm of 
Valiant that relies on Fourier-analytic 
techniques (cf.~\cite[Theorems 2.4, 5.2, and 5.6]{Valiant2015}).
Compared with Valiant's algorithm, our algorithm has a slightly worse 
tolerance for noise but better sample complexity:

\begin{restatable}{cor}{corsparseparity}
\label{cor:sparse-parity}
For all constants $0<\epsilon<\omega/3$, the \textsc{Parity with Noise}
problem admits a randomized algorithm that uses 
\begin{equation}
\label{eq:sparse-parity-examples}
d\geq (2k+3)\cdot|1-2\eta|^{-\frac{4\omega}{9\epsilon}-\frac{2}{3}}\log n
\end{equation}
examples, runs in time
\begin{equation}
\label{eq:sparse-parity-bound}
\tilde O\bigl(n^{\frac{\omega+\epsilon}{3}k}\cdot|1-2\eta|^{-\frac{8\omega}{9\epsilon}-\frac{4}{3}}\bigr)\,,
\end{equation}
for all sufficiently large $k$, and finds the support of the parity function, with probability %
at least $1-o(1)$.
The running time bound holds uniformly for all $n^{-\Theta(1)}<|1-2\eta|<1$.
\end{restatable}

Using algorithms of Feldman, Gopalan, Khot, and Ponnuswami~\cite{Feldman2009}, 
and Mossel, O'Donnell, and Servedio~\cite{Mossel2004}, 
from Corollary~\ref{cor:sparse-parity} one can obtain further corollaries in 
the context of learning sparse juntas and DNFs. We refer to 
\cite{Valiant2015} for a detailed exposition.

\subsection{Subquadratic Algorithms in Small Dimension}

If the dimension $d$ is very small, then subquadratic algorithms in $n$ 
for \textsc{Out\-lier Cor\-re\-la\-tions} are
available through dedicated space-partitioning data structures such as 
\textit{Voronoi diagrams}, enabling near neighbor query times 
$\tilde O(d^c)$ for a constant $c$, but with exponential scaling in size as
a function of $d$~\cite{Meiser1993,Weber1998}. Similar query-vs-size 
tradeoffs are available also in higher dimensions; for example, 
Kushilevitz, Ostrovsky, and Rabani~\cite{Kushilevitz1998} presented 
a data structure with $\tilde O(d)$ query time and size $(dn)^{O(1)}$
for approximate nearest neighbors.
Alman and Williams~\cite{Alman2015} obtain subquadratic scaling in $n$ for $d=O(\log n)$ 
(see \S\ref{sect:lower-upper}).

\subsection{Scaling in the Number of Pairs above the Background Correlation}

Let us briefly study scaling in the case when the parameter $q$ is 
relatively large, such as the {\em batch-query}%
\footnote{Suppose we have $n$ observables in a database
and $n$ observables that constitute queries to the database. 
Suppose furthermore that for each query there are $O(1)$ outlier-correlated 
observables in the database, and that our task is to find these outlier-correlated observables for each query.}{}
case with $q=O(n)$, or the case $q=O(n^{2-\delta})$ for 
a small constant $\delta>0$. 
In this situation we observe that in Theorem~\ref{thm:main} the term 
$\tilde O(qdn^{2\gamma})$ can dominate the running time over 
\eqref{eq:main-bound}. Indeed, assuming $qd$ is subquadratic in $n$, we can 
always obtain an overall running time that is subquadratic in $n$ by forcing 
a small enough $\gamma>0$ in Theorem~\ref{thm:main}. However, such a forced 
$\gamma$ may cause a suboptimal value of \eqref{eq:main-bound} over what 
would be available if $q$ was smaller. This bottleneck in scaling for 
large $q$ can be somewhat alleviated by pursuing a two-level recursive 
strategy, where we run the approximate detection algorithm recursively 
before proceeding to listing:

\begin{restatable}[Main, Two-level]{theorem}{thmmaintwolevel}
\label{thm:main-two-level}
For all constants $0<\gamma,\kappa<1$ and $\Delta\geq 1$,
the \textsc{Out\-lier Cor\-re\-la\-tions} problem for Boolean inputs
admits a randomized algorithm that runs in 
time 
\[
\tilde O\bigl(n^{\max\,\{1-\gamma+M(\Delta\gamma,\gamma),\,M(1-\gamma,2\Delta\gamma)\}}\tau^{-4}\bigr)
\]
for approximate detection and subsequent
\[
\tilde O\bigl(qn^{\gamma\max\,\{1-\kappa+M(\Delta\kappa,\kappa),\,M(1-\kappa,2\Delta\kappa)\}}\tau^{-4}+qdn^{2\gamma\kappa}\bigr)
\]
time for exact listing of all the outliers, w.h.p.
The running time bounds hold uniformly for 
all $n^{-\Theta(1)}\leq\tau<\rho<1$ with 
$\log_\tau\rho\leq 1-\Delta^{-1}$.
\end{restatable}

For specific choices $\gamma,\kappa,\Delta$ we obtain
the analogs of Corollary~\ref{cor:omega} and \ref{cor:alpha}.

\begin{restatable}{cor}{coromegatwolevel}
\label{cor:omega-two-level}
For all constants $0<\tau<\rho<1$, 
the \textsc{Out\-lier Cor\-re\-la\-tions} problem for Boolean inputs
admits a randomized algorithm that runs in time 
\[
\tilde O\left(n^{\frac{2\omega}{3-\log_\tau\rho}}\right)
\]
for approximate detection and subsequent 
\[
\tilde O\left(qn^{\frac{2\omega(1-\log_\tau\rho)}{(3-\log_\tau\rho)^2}}
             +qdn^{\frac{2(1-\log_\tau\rho)^2}{(3-\log_\tau\rho)^2}}\right)
\]
time for exact listing of all the outliers, w.h.p.
\end{restatable}

\begin{restatable}{cor}{coralphatwolevel}
\label{cor:alpha-two-level}
For all constants $0<\tau<\rho<1$, 
the \textsc{Out\-lier Cor\-re\-la\-tions} problem for Boolean inputs
admits a randomized algorithm that runs in time 
\[
\tilde O\left(n^{\frac{4}{2+\alpha(1-\log_\tau\rho)}}\right)
\]
for approximate detection and subsequent 
\[
\tilde O\left(qn^{\frac{4\alpha(1-\log_\tau\rho)}{(2+\alpha(1-\log_\tau\rho))^2}}+qdn^{\frac{2\alpha^2(1-\log_\tau\rho)^2}{(2+\alpha(1-\log_\tau\rho))^2}}\right)
\]
time for exact listing of all the outliers, w.h.p.
\end{restatable}

\begin{remark}
Let us recall from Remark~\ref{rem:weak} 
that Corollary~\ref{cor:alpha} lists outliers with 
$\rho=2^{-100}$ and $\tau=2^{-101}$ in time $\tilde O(n^{1.998}+qdn^{0.003})$.
Corollary~\ref{cor:alpha-two-level} improves this to
time $\tilde O(n^{1.998}+qn^{0.003}+qdn^{0.0000045})$.
\end{remark}

\begin{remark}
In the setting of well-separated outliers with any constant $\rho$
and $\tau\rightarrow 0$, Corollary~\ref{cor:omega} lists outliers
in time $\tilde O(n^{2\omega/3+\epsilon_\tau}+qdn^{2/3+\epsilon_\tau})$. 
Corollary~\ref{cor:omega-two-level} improves this to 
time $\tilde O(n^{2\omega/3+\epsilon_\tau}+qn^{2\omega/9+\epsilon_\tau}
+qdn^{2/9+\epsilon_\tau})$. Here $\epsilon_\tau\rightarrow 0$
as $\tau\rightarrow 0$.
\end{remark}

\smallskip

\subsection{Discussion and Further Work}

In addition to Valiant's algorithm, another 
breakthrough in the context of finding and approximating the outlier entries
in a matrix product $A^\top\!B$ of two given $t\times s$ matrices 
$A,B$ was made by 
Pagh~\cite{Pagh2013}. 
Pagh's algorithm also computes a 
{\em compressed version} of the product $A^\top\!B$ by using 2-wise
independent hash families that are algebraically compatible
with a cyclic group (of order $b$) to aggregate the operands $A,B$ 
with random signs into preimages of the hash function along the outer 
dimension ($s$) and with explicit (uncompressed) summation along the 
inner dimension ($t$). 
Because of the compatibility with the cyclic group, the compressed
operands can be multiplied in essentially linear time via cyclic convolution 
(using the FFT for the cyclic group). The compressed product gives 
unbiased estimates of the product entries $(A^\top\!B)_{j_1j_2}$ with 
2-wise independence controlling the variance via the
Frobenius norm at $\|A^\top\!B\|_F^2/b$.
Despite the essentially linear-time performance given by the use of FFT 
in evaluating the compressed products (thus enabling a large value of $b$),
the combination of Frobenius control on variance and the linear
scaling along the inner dimension $t$ appears not to yield improvements
for the light bulb problem or for \textsc{Out\-lier Cor\-re\-la\-tions}. 

To our knowledge, there are no lower bounds that would preclude
the use of multilinear algorithms other than fast matrix multiplication 
(such as Pagh's~\cite{Pagh2013} algorithm discussed above)
 for \textsc{Out\-lier Cor\-re\-la\-tions}, yet fast matrix multiplication 
remains the only known tool to obtain truly subquadratic scaling for 
weak outliers. As Valiant~\cite[\S6]{Valiant2015} highlights, 
it would be of considerable interest to avoid fast matrix multiplication 
altogether to obtain practical subquadratic scaling for 
moderately-sized $n$.
In fact, from this perspective our present results arguably proceed in the 
wrong direction by making heavier use of fast matrix multiplication 
to obtain asymptotically faster subquadratic scaling.
For the state of the art in practical fast matrix multiplication
algorithms, cf.~\cite{Huang2017} and \cite{Benson2015}.

Recently, Ahle, Pagh, Razenshteyn, and Silvestri~\cite{Ahle2015} 
show that inner product similarity joins are hard to approximate
in subquadratic time in $n$ unless the Orthogonal Vectors
Conjecture (see~\S\ref{sect:orthogonal-vectors}) and subsequently SETH
 are false. Stated in terms
of \textsc{Out\-lier Cor\-re\-la\-tions}, they show that Boolean inputs
with $\log_\tau\rho=1-o(1/\sqrt{\log n})$ do not admit subquadratic
scaling in $n$ unless OVC is false which would imply that SETH is false.
Narrowing the gap 
between subquadratic approximability and inapproximability remains 
a topic for further work. 

Let us conclude with a few questions for further work. 
Is $\log_\tau\rho$ the natural parameter for subquadratic scaling 
in the context of \textsc{Out\-lier Cor\-re\-la\-tions}? Is it possible to improve 
the limiting exponent $2\omega/3$ in Corollary~\ref{cor:omega} and in
Corollary~\ref{cor:light-bulb}? Curiously, the algorithm in our 
Corollary~\ref{cor:sparse-parity} and Valiant's algorithm~\cite[Algorithm~11]{Valiant2015}  
for \textsc{Parity with Noise} rely on somewhat different techniques, 
yet both arrive at the constant $\omega/3$ for learning sparse parities%
---is this just a coincidence? Finally, is it possible to derandomize our 
algorithms or improve their space usage? In subsequent work~\cite{Karppa2016},
a superset of present authors derandomize Valiant's algorithm, but restricted to
$(\pm 1)$-valued inputs only and with more modest subquadratic running times. 
We refer to \cite{Pagh2016} and \cite{Kapralov2015} for further recent work 
and motivation for derandomization and resource tradeoffs in the context of 
similarity search.


\section{Preliminaries}

This section collects terminology, notation, and background results 
used in the subsequent development.

\subsection{The Hoeffding Bound}

The Hoeffding bound establishes sharp concentration around 
the expectation of a sum of independent terms assuming we have 
control on the support of the terms.
\begin{theorem}[{Hoeffding~\cite[Theorem~2]{Hoeffding1963}}]
Let $Z_1,Z_2,\ldots,Z_s$ be independent random variables
with $\ell_i\leq Z_i\leq u_i$ for all $i=1,2,\ldots,s$
and let $Z=Z_1+Z_2+\ldots+Z_s$. Then, for all $d>0$ it holds that 
\begin{equation}
\label{eq:hoeffding}
\Pr\bigl(Z-\Ex\bigl[Z\bigr]\geq d\bigr)
  \leq \exp\biggl(-\frac{2d^2}{\sum_{i=1}^s(u_i-\ell_i)^2}\biggr)\,.
\end{equation}
\end{theorem}

\subsection{Inner Products, Restriction, Powering}

Let us set up some notation regarding inner products and tensor powers.
For convenience, let us write $[d]=\{1,2,\ldots,d\}$. 
Let $x=(x_1,x_2,\ldots,x_d)$ and $y=(y_1,y_2,\ldots,y_d)$ 
be two vectors of indeterminates. The inner product
\begin{equation}
\label{eq:inner}
\bra x,y\ket = x_1y_1+x_2y_2+\ldots+x_dy_d
\end{equation}
is a multilinear polynomial in the indeterminates $x,y$. 
For a (multi)set of indices \mbox{$I\subseteq[d]$}, we use the notation
\[
\bra x,y\ket_I=\sum_{i\in I}x_iy_i
\]
for an inner product $\bra x,y\ket$ taken over the indices in $I$.

For a positive integer $p$, we observe the $p$th 
power of the inner product $\bra x,y\ket$ is a polynomial in $x,y$ 
whose monomials 
can be indexed by $p$-tuples $\vec i=(i_1,i_2,\ldots,i_p)\in [d]^p$
with $i_1,i_2,\ldots,i_p\in[d]$.
Indeed, expanding the product of sums into a sum of products,
we observe that
\begin{equation}
\label{eq:power-poly}
\bra x,y\ket^p 
= (x_1y_1+x_2y_2+\ldots+x_dy_d)^p
= \sum_{\vec i\in[d]^p}\prod_{\ell=1}^p x_{i_\ell}\prod_{\ell=1}^p y_{i_\ell}\,.
\end{equation}

Let us write $x^{\otimes p}$ and $y^{\otimes p}$ for the
$d^p$-dimensional vectors obtained by taking the $p$th 
tensor power of $x$ and $y$ and whose coordinates are indexed 
by the $d^p$ possible $p$-tuples:
\[
x^{\otimes p}=\left(\prod_{\ell=1}^p x_{i_\ell}:\vec i\in [d]^p\right)
\text{ and } 
y^{\otimes p}=\left(\prod_{\ell=1}^p y_{i_\ell}:\vec i\in [d]^p\right)\,.
\]
With this notation, we observe from \eqref{eq:inner} and 
\eqref{eq:power-poly} that
\begin{equation}
\label{eq:power-inner}
\bra x,y\ket^p 
=\bra x^{\otimes p},y^{\otimes p}\ket\,.
\end{equation}

\subsection{Cartesian Sampling Lemma}
The following lemma will be convenient when analysing the concentration
of samples obtained in the compression algorithm. For an analogous
analysis, cf.~\cite{DKM2006}.

\begin{lemma}[Cartesian Sampling]
\label{lem:cartesian}
Let $s$ and $m$ be positive integer squares and let 
$x_1,x_2,\ldots,x_{m^{1/2}}\in\{-1,1\}$
and
$y_1,y_2,\ldots,y_{m^{1/2}}\in\{-1,1\}$
with 
\[
\sum_{u=1}^{m^{1/2}}x_u=\xi m^{1/2}\qquad\text{and}\qquad
\sum_{v=1}^{m^{1/2}}y_v=\eta m^{1/2}
\]
for some $-1\leq\xi,\eta\leq 1$.
Suppose $S_1,S_2$ are two multisets of size $s^{1/2}$ 
selected by drawing two $s^{1/2}$-tuples consisting of 
elements of $[m^{1/2}]$ independently and uniformly at random 
(with repetition). 
Then, with high probability in $s$, we have
\begin{equation}
\label{eq:cartesian-upper}
\biggl|\sum_{(u,v)\in S_1\times S_2}x_uy_v\biggr|
\leq|\xi\eta|s +(|\xi|+|\eta|)s^{3/4}\log s+s^{1/2}(\log s)^2\,.
\end{equation}
Furthermore, if 
\begin{equation}
\label{eq:cartesian-xi-eta}
|\xi|,|\eta|\geq s^{-1/4}\log s\,, 
\end{equation}
then, with high probability in $s$, we have
\begin{equation}
\label{eq:cartesian-lower}
\biggl|\sum_{(u,v)\in S_1\times S_2}x_uy_v\biggr|
\geq
|\xi\eta|s-(|\xi|+|\eta|)s^{3/4}\log s+s^{1/2}(\log s)^2\,.
\end{equation}
\end{lemma}

\begin{proof}
For $j=1,2,\ldots,s^{1/2}$ let 
$X_j\in\{-1,1\}$ be a random variable that independently assumes each of
the $m^{1/2}$ values $x_u$ with probability $m^{-1/2}$. 
For $j=1,2,\ldots,s^{1/2}$ let 
$Y_j\in\{-1,1\}$ be a random variable that independently assumes each of
the $m^{1/2}$ values $y_v$ with probability $m^{-1/2}$. 
Consider the random variables $X=X_1+X_2+\ldots+X_{s^{1/2}}$
and $Y=Y_1+Y_2+\ldots+Y_{s^{1/2}}$. 
By linearity of expectation, we have
$\Ex[X]=\xi s^{1/2}$ and $\Ex[Y]=\eta s^{1/2}$.
Furthermore, from the Hoeffding bound \eqref{eq:hoeffding}
it follows that
$|X-\Ex[X]|\leq s^{1/4}\log s$ 
and
$|Y-\Ex[Y]|\leq s^{1/4}\log s$ with high probability in $s$.
Thus, we conclude that with high probability in $s$ it holds that
\begin{align*}
|XY| & \leq \bigl(|\xi|s^{1/2}+s^{1/4}\log s\bigr) \bigl(|\eta|s^{1/2}+s^{1/4}\log s\bigr) \\
& = |\xi\eta|s+s^{3/4}(|\xi|+|\eta|)\log s+s^{1/2}(\log s)^2.
\end{align*}
Similarly, assuming that $|\xi|,|\eta|\geq s^{-1/4}\log s$,
with high probability in $s$ we have the lower bound
\[
\begin{split}
|XY| &
\geq 
\bigl(|\xi|s^{1/2}-s^{1/4}\log s\bigr)
\bigl(|\eta|s^{1/2}-s^{1/4}\log s\bigr)\\
&=
|\xi\eta|s
-s^{3/4}(|\xi|+|\eta|)\log s
+s^{1/2}(\log s)^2\,.
\end{split}
\]
\end{proof}

\subsection{Anti-concentration of Signed Aggregation}

We recall the following anti-concentration lemma from
the analysis of Valiant's \mbox{algorithm}.

\begin{lemma}[{Valiant~\cite[Lemma~3.2]{Valiant2015}}]
\label{lem:anti-concentration}
Let $C$ be a $t\times t$ matrix with entries $c_{ij}$.
Suppose that $\alpha_1,\alpha_2,\ldots,\alpha_t\in\{-1,1\}$
and $\beta_1,\beta_2,\ldots,\beta_t\in\{-1,1\}$ have
been selected independently and uniformly at random.
Then, 
\[
\Pr\biggl(\biggl|\sum_{i,j} \alpha_i\beta_jc_{ij}\biggr|\geq\frac{1}{4}\max_{i,j}|c_{ij}|\biggr)\geq\frac{1}{4}\,.
\]
\end{lemma}

\section{The Algorithm}

\label{sect:algorithm}

This section proves Theorem~\ref{thm:main}. We present a self-contained
analysis and description of the entire algorithm underlying
Theorem~\ref{thm:main} even if our main technical contribution occurs in 
the compression subroutine that simultaneously expands and aggregates
(cf.~\S\ref{sect:valiant-algorithm} and~\S\ref{sect:contribution}).%

\subsection{The Input and the Parameters}

Let $0<\gamma<1$ and $\Delta\geq 1$ be fixed constants. 
Suppose we are given as input two $d\times n$ matrices 
$A,B\in\{-1,1\}^{d\times n}$ together with parameters $\rho,\tau$ 
that satisfy $n^{-\Theta(1)}\leq\tau<\rho<1$ and 
$\log_\tau\rho\leq 1-\Delta^{-1}$. 
For $i\in [d]$ and $j\in [n]$, let us write $a_{ij}$ and $b_{ij}$ for 
the entries of $A$ and $B$ at row $i$, column $j$. 
Furthermore, let us write $a_j$ and $b_j$ for the $j$th column of $A$ and $B$,
respectively. 

The algorithm works with three positive integer parameters $s,t,p$ 
whose precise values we will fix in what follows.
At this point, we can assume that $p$ is even, 
and that $s$ is a positive integer square. 
We will furthermore pad our input matrices with at most $t-1 < n$
all-zero columns to ensure divisibility by $t$. Let us denote 
by $\bar n = \lceil n/t \rceil \cdot t$ the number of columns including padding.

\subsection{The Algorithm}

We describe the algorithm first and then proceed to analyse its
correctness and running time. The algorithm executes 
a five-phase iteration $\lceil(\log n)^2\rceil$ times, 
and gives as its output the union of the outputs of all the iterations.
The five phases are as follows:

{\em 0. Setup.}  Draw uniformly at random a partition
$J_1,J_2,\ldots,J_{\bar n/t}$ of $[\bar n]$ into sets $J_k$ with
$|J_k|=t$ for all $k\in[\bar n/t]$.  For $k\in [\bar n/t]$, let us write
$\underline J_k = J_k \cap [n]$ for the restriction of $J_k$ to
the indices corresponding to non-zero columns. We observe that
$1 \leq |\underline J_k| \leq t$.
Draw independently and uniformly at random (with repetition)
two $s^{1/2}$-tuples consisting of elements of $[d]^{p/2}$ 
and let $I_1$ and $I_2$ be the resulting multisets
of size $s^{1/2}$. Form the Cartesian product $I=I_1\times I_2$ of size $s$
and observe that all the elements of $I$ are $p$-tuples in $[d]^p$. 
Draw $\alpha_1,\alpha_2,\ldots,\alpha_{\bar n}\in\{-1,1\}$ 
and $\beta_1,\beta_2,\ldots,\beta_{\bar n}\in\{-1,1\}$ 
independently and uniformly at random.

{\em The compressed matrices (definition).}
Let us first define what we compute and only 
then give the algorithm that computes the matrices.
For $\vec i\in I$ and $k\in[\bar n/t]$, define 
\begin{equation}
\label{eq:hat}
\begin{aligned}
\hat a_{\vec ik}=\sum_{j\in \underline J_k}\alpha_j(a_j^{\otimes p})_{\vec i} 
\qquad\text{and} \qquad \hat b_{\vec ik}=\sum_{j\in \underline J_k}\beta_j(b_j^{\otimes p})_{\vec i}\,.
\end{aligned}
\end{equation}
This defines the matrices $\hat A_I=(\hat a_{\vec ik})$ 
and $\hat B_I=(\hat a_{\vec ik})$, both of size $s\times (\bar n/t)$.%
\footnote{%
We observe in particular these matrices are essentially what results if we 
execute phases and 1 and 2 of Valiant's algorithm 
(recall~\S\ref{sect:valiant-algorithm}), with the difference that
our sample $I$ always has Cartesian product structure, whereas
Valiant's algorithm uses a uniform random sample of $[d]^p$.}

{\em 1+2. Compression by simultaneous expansion and aggregation.}
We compute the compressed matrices $\hat A_I$ and $\hat B_I$ from the 
given input $A$ and $B$ as follows.
We describe the computation only for $\hat A_I$, the computation 
for $\hat B_I$ is symmetric.
Let us first establish that the task of computing $\hat A_I$ amounts to
$\bar n/t$ matrix products {\em because our random sample $I$
decomposes into the Cartesian product $I=I_1\times I_2$}.
Indeed, we observe from \eqref{eq:hat} that 
if we write $\vec i=(\vec i_1,\vec i_2)\in I$ in terms of its parts
$\vec i_1\in I_1$ and $\vec i_2\in I_2$, we have 
\begin{equation}
\label{eq:hat-cartesian}
\hat a_{\vec ik}=
\hat a_{(\vec i_1,\vec i_2),k}=
\sum_{j\in \underline J_k}\alpha_j(a_j^{\otimes p/2})_{\vec i_1}
                       (a_j^{\otimes p/2})_{\vec i_2}\,.
\end{equation}
Thus, for each fixed $k\in[\bar n/t]$ we observe that
\eqref{eq:hat-cartesian} is in fact a matrix product. 
Indeed, define the $s^{1/2}\times t$ matrix 
$L_k=(\ell_{\vec i_1j})$ to consist of the entries 
\begin{equation}
\label{eq:lk}
\ell_{\vec i_1j}=\alpha_j(a_j^{\otimes p/2})_{\vec i_1}
\end{equation}
for $\vec i_1\in I_1$ and $j\in J_k$. Similarly, 
define the $s^{1/2}\times t$ matrix 
$R_k=(r_{\vec i_2j})$ to consist of the entries 
\begin{equation}
\label{eq:rk}
r_{\vec i_2j}=(a_j^{\otimes p/2})_{\vec i_2}
\end{equation}
for $\vec i_2\in I_2$ and $j\in J_k$. 
By \eqref{eq:hat} the $(\vec i_1,\vec i_2)$-entry of 
the product matrix $L_kR_k^\top $ is precisely the value 
$\hat a_{(\vec i_1,\vec i_2),k}$. Indeed,
\[
(L_kR_k^\top )_{\vec i_1,\vec i_2}
=\sum_{j\in J_k}\ell_{\vec i_1j}r_{\vec i_2j}
=\sum_{j\in \underline J_k}\ell_{\vec i_1j}r_{\vec i_2j}
=\sum_{j\in \underline J_k}\alpha_j(a_j^{\otimes p/2})_{\vec i_1}
                        (a_j^{\otimes p/2})_{\vec i_2}
=\hat a_{(\vec i_1,i_2),k}\,.
\]
To compute the matrix $\hat A_I$, we thus proceed as follows. 
For each $k\in [\bar n/t]$, we
construct the matrices $L_k$ and $R_k$ entrywise using \eqref{eq:lk}
and \eqref{eq:rk}, respectively, and then we compute the
$s^{1/2}\times s^{1/2}$ product $L_kR_k^\top$. 
This gives us the $(\bar n/t)\times s$ matrix $\hat A_I$, 
at the cost of $\bar n/t$ matrix multiplications with 
inner dimension $t$ and outer dimension $s^{1/2}$.

{\em 3. Approximate detection.}
We compute the matrix product $\hat A_I^\top \hat B_I$
and {\em mark} all entries $(k_1,k_2)\in [\bar n/t]\times [\bar n/t]$
that satisfy $|(\hat A_I^\top \hat B_I)_{k_1k_2}|\geq\frac{1}{8}\rho^ps$. 
(This requires one matrix product with inner dimension $s$ and 
outer dimension $\bar n/t$.)

{\em 4. Exact listing.}
Finally, for each marked entry $(k_1,k_2)\in [\bar n/t]\times [\bar n/t]$,
we iterate over all $(j_1,j_2)\in \underline J_{k_1}\times \underline J_{k_2}$,
compute the inner product $\bra a_{j_1},b_{j_2}\ket$, and
output $(j_1,j_2)$ as an outlier pair 
if $|\bra a_{j_1},b_{j_2}\ket|\geq \rho d$.

\subsection{Analysis (correctness)}
\label{sect:correctness}

\emph{Proof outline.} The correctness proof proceeds as follows. We
start by establishing an upper bound on the absolute value of
entries in the product matrix $\hat A_I^\top \hat B_I$,
assuming only small partial inner products are aggregated in an entry. 
Dually, we then establish a lower bound for an entry of 
$\hat A_I^\top \hat B_I$ that contains at least one large partial inner 
product. We then establish control on the concentration of the 
partial inner products using 
the Cartesian concentration lemma (Lemma~\ref{lem:cartesian}).
With appropriate parameterization, we then verify that any single outlier 
is found (its entry in the product matrix is marked for listing) 
with good probability during one iteration, and amplify this probability 
by iteration to ensure that all outliers are found with high probability. 
Finally, we show that at most $q$ entries are marked for listing 
with high probability, completing the correctness proof.

Let us now proceed with the detailed proof. 
We start by fixing the value of the parameter $t$. 
With foresight, let $t$ be the unique integer that satisfies
\begin{equation}
\label{eq:t}
n^{\gamma}\leq t<n^{\gamma}+1\,. 
\end{equation}
Observe that the choice implies $n\leq \bar n\leq n+n^\gamma$.

We begin our analysis by studying the entries of the
product matrix $\hat A_I^\top \hat B_I$. Let $k_1,k_2\in [\bar n/t]$ and 
consider the $(k_1,k_2)$-entry of $\hat A_I^\top \hat B_I$.
From \eqref{eq:hat} we have 
\begin{equation}
\label{eq:output-entry}
\begin{split}
\bigl(\hat A_I^\top \hat B_I\bigr)_{k_1k_2}
&=\sum_{\vec i\in I}\biggl(
  \sum_{j_1\in \underline J_{k_1}}\alpha_{j_1}\bigl(a_{j_1}^{\otimes p}\bigr)_{\vec i}\biggr)
 \biggl(
  \sum_{j_2\in \underline J_{k_2}}\beta_{j_2}\bigl(b_{j_2}^{\otimes p}\bigr)_{\vec i}\biggr)\\
&=\sum_{j_1\in \underline J_{k_1}} \alpha_{j_1}
  \sum_{j_2\in \underline J_{k_2}} 
  \beta_{j_2} \bra a_{j_1}^{\otimes p},b_{j_2}^{\otimes p}\ket_I\,.
\end{split}
\end{equation}

{\em The value of an entry of $\hat A_I^\top \hat B_I$ that aggregates only small partial inner products.}
Let us first derive an upper bound for $|\bigl(\hat A_I^\top \hat B_I\bigr)_{k_1k_2}|$ subject to an upper bound for the absolute values of the partial 
inner products $|\bra a_{j_1}^{\otimes p},b_{j_2}^{\otimes p}\ket_I|$.
That is, let us assume $U\geq 0$ is an upper bound 
$|\bra a_{j_1}^{\otimes p},b_{j_2}^{\otimes p}\ket_I|\leq U$ 
that holds for all $j_1\in J_{k_1}$ and $j_2\in J_{k_2}$
with high probability in $n$.
Recall that $|J_{k_1}|=|J_{k_2}|=t$ and that the signs
$\alpha_{j_1},\beta_{j_2}\in\{-1,1\}$ have been selected independently
and uniformly at random for $j_1\in J_{k_1}$ and $j_2\in J_{k_2}$.
Let us fix $j_1\in J_{k_1}$ arbitrarily and analyse the concentration of 
the innermost sum $Z_{j_1}=\sum_{j_2\in  J_{k_2}} \beta_{j_2} \bra a_{j_1}^{\otimes p},b_{j_2}^{\otimes p}\ket_I$ in \eqref{eq:output-entry}.
Because $\beta_{j_2}$ and $\bra a_{j_1}^{\otimes p},b_{j_2}^{\otimes p}\ket_I$
are independent, each summand $Z_{j_1,j_2}=\beta_{j_2} \bra a_{j_1}^{\otimes p},b_{j_2}^{\otimes p}\ket_I$ is a random variable with zero expectation. 
Let us condition on our assumption 
$|\bra a_{j_1}^{\otimes p},b_{j_2}^{\otimes p}\ket_I|\leq U$
and thus conclude that $|Z_{j_1,j_2}|\leq U$ holds for each summand. 
Thus, since $t\leq n$, the Hoeffding bound 
\eqref{eq:hoeffding} gives us that $|Z_{j_1}|\leq t^{1/2}U\log n$ 
with high probability in $n$.
By the union bound we thus have that 
$|Z_{j_1}|\leq t^{1/2}U\log n$ holds {\em for all $j_1\in J_{k_1}$}
with high probability in $n$.
Let us condition on this event.
Observe that \eqref{eq:output-entry} equals 
$Z=\sum_{j_1\in J_{k_1}}\alpha_{j_1}Z_{j_1}$.
Since $\alpha_{j_1}$ and $Z_{j_1}$ are independent, 
the product $\alpha_{j_1}Z_{j_1}$ has zero expectation. 
Thus, from $|\alpha_{j_1}Z_{j_1}|\leq t^{1/2}U\log n$
and the Hoeffding bound \eqref{eq:hoeffding},
we conclude that, with high probability in $n$, we have
\begin{equation}
\label{eq:outer-upper}
\left|\bigl(\hat A_I^\top \hat B_I\bigr)_{k_1k_2}\right|\leq tU(\log n)^2\,.
\end{equation}

{\em The value of an entry of $\hat A_I^\top \hat B_I$ that aggregates at least one large partial inner product.}
Next, let us suppose that $L\geq 0$ is a lower bound 
$|\bra a_{j_1}^{\otimes p},b_{j_2}^{\otimes p}\ket_I|\geq L$
for some specific $j_1,j_2\in[n]$. 
Then, for the pair of blocks $(k_1,k_2)\in[\bar n/t]\times[\bar n/t]$ 
with $j_1\in \underline J_{k_1}$ and $j_2\in \underline J_{k_2}$
we conclude from \eqref{eq:output-entry} and Lemma~\ref{lem:anti-concentration}
that, with probability at least $1/4$, we have
\begin{equation}
\label{eq:outer-lower}
\left|\bigl(\hat A_I^\top \hat B_I\bigr)_{k_1k_2}\right|\geq 
\frac{L}{4}\,.
\end{equation}
When \eqref{eq:outer-lower} holds, we say the 
pair $(j_1,j_2)$ is {\em signalled by} (the $(k_1,k_2)$-entry in) 
$\hat A_I^\top \hat B_I$.

{\em Concentration of the partial inner products.}
Let $j_1,j_2\in [n]$. Let us now study the concentration of 
$\bra a_{j_1}^{\otimes p},b_{j_2}^{\otimes p}\ket_I$ using
the Cartesian concentration lemma (Lemma~\ref{lem:cartesian}). 
Because $I=I_1\times I_2$, the partial inner product 
$\bra a_{j_1}^{\otimes p},b_{j_2}^{\otimes p}\ket_{I}$ decomposes into
\begin{equation}
\label{eq:product-sum-form}
\begin{split}
\bra a_{j_1}^{\otimes p},b_{j_2}^{\otimes p}\ket_{I_1\times I_2}
&=\sum_{\vec i\in I_1\times I_2}
   \bigl(a_{j_1}^{\otimes p}\bigr)_{\vec i}
   \bigl(b_{j_2}^{\otimes p}\bigr)_{\vec i}\\
&=\sum_{(\vec i_1,\vec i_2)\in I_1\times I_2}\ \ 
   \bigl(a_{j_1}^{\otimes p/2}\bigr)_{\vec i_1}
   \bigl(a_{j_1}^{\otimes p/2}\bigr)_{\vec i_2}
   \bigl(b_{j_2}^{\otimes p/2}\bigr)_{\vec i_1}
   \bigl(b_{j_2}^{\otimes p/2}\bigr)_{\vec i_2}
\,.
\end{split}
\end{equation}
Now suppose that $\bra a_{j_1},b_{j_2}\ket= d\sigma$ for some 
$-1\leq\sigma\leq 1$. 
From \eqref{eq:power-inner} we thus have 
$\bra a_{j_1}^{\otimes p/2},b_{j_2}^{\otimes p/2}\ket=d^{p/2}\sigma^{p/2}$.
Recalling that $|I_1|=|I_2|=s^{1/2}$, 
take $\xi=\eta=\sigma^{p/2}$, $m=d^p$,
$S_1=I_1$, $S_2=I_2$, 
$x_{\vec u}=
   (a_{j_1}^{\otimes p/2}\bigr)_{\vec u}
   (b_{j_2}^{\otimes p/2}\bigr)_{\vec u}$,
and 
$y_{\vec v}=
   (a_{j_1}^{\otimes p/2}\bigr)_{\vec v}
   (b_{j_2}^{\otimes p/2}\bigr)_{\vec v}$
for all $\vec u,\vec v\in[d]^{p/2}$
in Lemma~\ref{lem:cartesian} and thus
observe from \eqref{eq:product-sum-form} and \eqref{eq:cartesian-upper} 
that, with high probability in $s$,
we have the upper bound
\begin{equation}
\label{eq:inner-upper}
\begin{split}
\bigl|\bra a_{j_1}^{\otimes p},b_{j_2}^{\otimes p}\ket_{I_1\times I_2}\bigr|
&=\biggl|\sum_{(\vec i_1,\vec i_2)\in I_1\times I_2}x_{\vec i_1}y_{\vec i_2}\biggr| \\
& \leq |\sigma^p|s
+2|\sigma^{p/2}|s^{3/4}\log s
+s^{1/2}(\log s)^2\,.
\end{split}
\end{equation}
Conversely, assuming that 
\begin{equation}
\label{eq:sigma-s}
|\sigma^p|\geq s^{-1/2}(\log s)^2
\end{equation}
so that \eqref{eq:cartesian-xi-eta} holds,
from \eqref{eq:product-sum-form} and \eqref{eq:cartesian-lower} we have,
with high probability in $s$, the lower bound
\begin{equation}
\label{eq:inner-lower}
\begin{split}
\bigl|\bra a_{j_1}^{\otimes p},b_{j_2}^{\otimes p}\ket_{I_1\times I_2}\bigr|
&\geq |\sigma^p|s
-2|\sigma^{p/2}|s^{3/4}\log s
+s^{1/2}(\log s)^2\\
&\geq |\sigma^p|s-2|\sigma^{p/2}|s^{3/4}\log s\,.
\end{split}
\end{equation}
Our eventual choice of $s$ will grow at least as fast as a root
function\footnote{By a root function of $n$ we mean a function $n^\beta$ for a constant $0<\beta<1$.}
of $n$, so by the union bound we can assume that the upper bound
\eqref{eq:inner-upper} and the lower bound \eqref{eq:inner-lower}
hold for all relevant $j_1,j_2\in [n]$ with high probability in $n$.

{\em An upper bound for aggregated background correlations.}
Let us now fix $s$ to be the least integer square that satisfies%
\footnote{Observe that for all large enough positive real $x$ 
the interval $[x,2x]$ contains at least one integer square.
Also observe that we have not yet fixed our value of $p$,
but will do so later.}
\begin{equation}
\label{eq:s}
\tau^{-2p}\leq s\leq 2\tau^{-2p}\,.
\end{equation}
Thus, from \eqref{eq:inner-upper} with $|\sigma|\leq\tau$ we have%
\begin{equation}
\label{eq:inner-upper-tau}
\begin{split}
\bigl|\bra a_{j_1}^{\otimes p},b_{j_2}^{\otimes p}\ket_{I_1\times I_2}\bigr|
&\leq 4\tau^p s(\log s)^2
\leq 4\tau^p s(\log n)^3\,;
\end{split}
\end{equation}
here we assume that our eventual choice for $s$ grows no faster than 
a polynomial function of $n$, that is $\log s=O(\log n)$ and hence 
$\log s\leq (\log n)^{3/2}$ for all large enough $n$.
Let us now choose a value for the upper bound $U$ in \eqref{eq:outer-upper}.
By \eqref{eq:inner-upper-tau} we can select 
$U=8\tau^p s(\log n)^3$
to conclude that, with high probability $n$,
for all $k_1,k_2\in[\bar n/t]$ we have
\begin{equation}
\label{eq:no-marking}
\left|\bigl(\hat A_I^\top \hat B_I\bigr)_{k_1k_2}\right|\leq 8\tau^p st(\log n)^5
\end{equation}
{\em unless} there is at least one pair 
$(j_1,j_2)\in \underline J_{k_1}\times \underline J_{k_2}$ with $|\bra a_{j_1},b_{j_2}\ket|>\tau d$.

{\em Any fixed outlier correlation will be listed with probability 
at least $1/4$ during any fixed iteration of the algorithm.}
For any outlier correlation we have $|\sigma|\geq\rho$. 
Let us assume that our eventual choice of $p$ will be such that for 
all large enough $p$ 
we have both
\begin{equation}
\label{eq:large-p-1}
\rho^p\geq \tau^p(\log s)^2\geq s^{-1/2}(\log s)^2
\end{equation}
and
\begin{equation}
\label{eq:large-p-2}
\ \ 2|\sigma^{p/2}|s^{3/4}\log s
\leq 2|\sigma^{p/2}|\tau^{p/2}s\log s
\leq\frac{1}{2}|\sigma^p| s\,.
\end{equation}
Subject to these assumptions, we observe that
\eqref{eq:sigma-s} holds by \eqref{eq:large-p-1} 
and from \eqref{eq:inner-lower} and \eqref{eq:large-p-2} we have
\begin{equation}
\label{eq:inner-lower-rho}
\bigl|\bra a_{j_1}^{\otimes p},b_{j_2}^{\otimes p}\ket_{I_1\times I_2}\bigr|
\geq \frac{1}{2}\rho^p s\,.
\end{equation}
Let us now choose a value for the lower bound $L$ in \eqref{eq:outer-lower}.
By \eqref{eq:inner-lower-rho} we can select
\hbox{$L=\frac{1}{2}\rho^p s$}
to conclude that any fixed $(j_1,j_2)\in[n]\times[n]$ 
with $|\bra a_{j_1},b_{j_2}\ket|\geq\rho d$ is signalled by 
$\hat A_I^\top \hat B_I$ with probability at least $1/4$.
Recall now that our threshold in the algorithm for marking an entry 
$(k_1,k_2)\in[\bar n/t]\times[\bar n/t]$ is 
$\left|(\hat A_I^\top \hat B_I)_{k_1k_2}\right|\geq\frac{1}{8}\rho^ps=L/4$.
Thus, assuming $(j_1,j_2)\in[n]\times[n]$ is 
signalled by $\hat A_I^\top \hat B_I$, the pair of blocks
$(k_1,k_2)\in [\bar n/t]\times[\bar n/t]$ with $(j_1,j_2)\in \underline J_{k_1}\times \underline J_{k_2}$ 
will be marked by the algorithm in the detection phase, and hence $(j_1,j_2)$ 
will be listed as an outlier in the listing phase.

{\em With high probability in $n$, all outlier correlations will be listed.}
Since the algorithm has at least $(\log n)^2$ independent
iterations, and each iteration outputs any fixed outlier correlation
with probability at least $1/4$, it follows that this fixed outlier
correlation is output with high probability in $n$. Taking the union
bound over the at most \mbox{$q\leq n^2$} outlier correlations, every outlier
correlation is listed with high probability in $n$. Observe that this
is unaffected by the padding of input data.

{\em With high probability in $n$, at most $q$ pairs of blocks will be marked.} 
It remains to complete the parameterization of the algorithm so that
at most $q$ pairs of blocks $(k_1,k_2)\in[\bar n/t]\times [\bar n/t]$ will be marked
with high probability in $n$.
Recall that there are at most $q$ pairs $(j_1,j_2)\in[n]\times[n]$
with $|\bra a_{j_1},b_{j_2}\ket|>\tau d$. Thus, recalling 
\eqref{eq:no-marking} and our marking threshold $\frac{1}{8}\rho^p s$,
it suffices to set up the parameters so that 
$8\tau^p st(\log n)^5 < \frac{1}{8}\rho^p s\,.$
For example, we can select the unique even integer $p$ with
\begin{equation}
\label{eq:p}
\begin{split}
&\frac{\log t+5\log \log n+\log 128}{\log(\rho/\tau)}\leq 
p
<
\frac{\log t+5\log \log n+\log 128}{\log(\rho/\tau)}+2
\,.
\end{split}
\end{equation}
From \eqref{eq:s} and \eqref{eq:p} we thus have
\begin{equation}
\label{eq:s-bound}
\begin{split}
&\ \ t^{\frac{2\log(1/\tau)}{\log(\rho/\tau)}}
(128\log n)^{\frac{10\log(1/\tau)}{\log(\rho/\tau)}}
\leq
s< 2t^{\frac{2\log(1/\tau)}{\log(\rho/\tau)}}
(128\log n)^{\frac{10\log(1/\tau)}{\log(\rho/\tau)}}\tau^{-4}\,.
\end{split}
\end{equation}
Now recall from \eqref{eq:t} that $n^{\gamma}\leq t<2n^\gamma$ and 
observe that 
\[
1\leq
\frac{\log(1/\tau)}{\log(\rho/\tau)}
=\frac{1}{1-\log_\tau \rho}\leq\Delta\,.
\]
Thus, from \eqref{eq:s-bound} we get
\begin{equation}
\label{eq:s-theta-n}
\begin{split}
&n^{2\gamma}(128\log n)^{10}
\leq 
s< 
2^{1+2\Delta}n^{2\Delta\gamma}(128\log n)^{10\Delta}\tau^{-4}\,.
\end{split}
\end{equation}
Since $\Delta\geq 1$ and $0<\gamma<1$ are constants, 
and recalling our assumption $\tau\geq n^{-\Theta(1)}$,
both $s$ and $t$ grow at least as fast as a root function of $n$
and at most as fast as a polynomial function of $n$.
Thus, high probabilities in $s$ and $t$ are high probabilities in $n$, 
and vice versa. It remains to justify our assumptions \eqref{eq:large-p-1} 
and \eqref{eq:large-p-2}. To establish \eqref{eq:large-p-1}, observe
that \eqref{eq:p} and $t\geq n^\gamma$ imply $(\rho/\tau)^p\geq n^\gamma$. 
Furthermore, \eqref{eq:s-theta-n} and $\tau\geq n^{-\Theta(1)}$
imply $(\log s)^2=O((\log n)^2)$. Thus, \eqref{eq:large-p-1} holds
by \eqref{eq:s}. To establish \eqref{eq:large-p-2}, recall that 
$|\sigma|\geq\rho$ and hence it suffices to establish 
$(\rho/\tau)^{p/2}\geq 4\log s$. The reasoning is similar to what
we used to establish \eqref{eq:large-p-1}.
This completes the correctness proof.

\subsection{Analysis (running time)}
\label{sect:running}

We recall Theorem~\ref{thm:main}:

\mainthm*

\begin{proof}
From \eqref{eq:s-theta-n} and \eqref{eq:t} we have 
$s=\tilde O(n^{2\Delta\gamma}\tau^{-4})$ and $t=\Theta(n^\gamma)$.
The matrices $\hat A_I$ and $\hat B_I$ are computed with 
$\bar n/t$ matrix multiplications, each with outer dimension 
$s^{1/2}=\tilde O(n^{\Delta\gamma}\tau^{-2})$ and inner dimension 
$t=\Theta(n^\gamma)$. The exponent of running time for computing 
the matrices $\hat A_I$ and $\hat B_I$ from the input $A$ and $B$ is thus 
\[
1-\gamma+M(\Delta\gamma-2\log_n\tau,\gamma)
\leq 1-\gamma+M(\Delta\gamma,\gamma)-4\log_n\tau\,.
\]
The matrix product $\hat A_I^\top \hat B_I$ has outer dimension $\bar n/t$ 
and inner dimension $s$. Thus, the exponent of running time 
for multiplying $\hat A_I^\top$ and $\hat B_I$ is 
\[
M(1-\gamma,2\Delta\gamma-4\log_n\tau)
\leq M(1-\gamma,2\Delta\gamma)-4\log_n\tau\,.
\]
The running time \eqref{eq:main-bound} for approximate detection 
in Theorem~\ref{thm:main} thus follows. (Indeed, the 
$\tilde O(\cdot)$-notation subsumes the polylogarithmic factors resulting 
from iterating the algorithm.) 

To obtain the the running time for listing, by the analysis in 
\S\ref{sect:correctness} at most $q$ entries of $\hat A_I^\top \hat B_I$ are marked
with high probability in $n$. Each marked
entry induces a computation of $t^2=\Theta(n^{2\gamma})$ inner products of 
dimension $d$, which results in the claimed running time
$\tilde O(qdn^{2\gamma})$.
This completes the proof of Theorem~\ref{thm:main}.
\end{proof}

\section{Corollaries}

This section proves the corollaries in \S\ref{sect:introduction} 
and \S\ref{sect:related-work}.

\label{sect:corollaries}

\subsection{Proof of Corollary~\ref{cor:omega}}
\label{sect:cor-omega}

We recall Corollary~\ref{cor:omega}:

\coromega*

\begin{proof}
Let us recall the following basic property of matrix multiplication
exponents. We have
\begin{equation}
\label{eq:mm-exp}
M(\mu,\nu)
\leq
\begin{cases}
(\omega-1)\mu+\nu  & \text{if $\mu\leq\nu$;}\\
2\mu+(\omega-2)\nu & \text{if $\mu > \nu$}.
\end{cases}
\end{equation}
Let us now parameterize Theorem~\ref{thm:main} to obtain 
Corollary~\ref{cor:omega}. Let us take 
$\gamma=\frac{1}{2\Delta+1}\,.$
In this case we have
$1-\gamma = 2\Delta\gamma\,.$
Recall the two terms in the maximum in \eqref{eq:main-bound}.
Since $\Delta\geq 1$, from \eqref{eq:mm-exp} we have
that the first term of the maximum in \eqref{eq:main-bound}
is bounded by 
\begin{equation}
\label{eq:e1-cor1}
1-\gamma+M(\Delta\gamma,\gamma)\leq\frac{4\Delta-2+\omega}{2\Delta+1}\,.
\end{equation}
The second term of the maximum in \eqref{eq:main-bound} is
\begin{equation}
\label{eq:e2-cor1}
M(1-\gamma,2\Delta\gamma)=\frac{2\Delta\omega}{2\Delta+1}\,.
\end{equation}
In particular, for $2\leq\omega\leq 3$ and $\Delta\geq 1$ 
we observe that $2\Delta\omega\geq 4\Delta-2+\omega$,
thus \eqref{eq:e2-cor1} dominates \eqref{eq:e1-cor1}.
Let us take 
$\Delta=\frac{1}{1-\log_\tau\rho}$
and observe that \eqref{eq:e2-cor1} simplifies to
$\frac{2\omega}{3-\log_\tau\rho}\,.$
This establishes the running time for approximate detection 
in Corollary~\ref{cor:omega}. The running time for listing is 
immediate by our choice of $\gamma$.
\end{proof}

\subsection{Proof of Corollary~\ref{cor:alpha}}
\label{sect:cor-alpha}

We recall Corollary~\ref{cor:alpha}:

\coralpha*

\begin{proof}
Let us take
$\gamma=\frac{\alpha}{2\Delta+\alpha}\,,$
where $0.30298<\alpha\leq 1$ is the exponent for rectangular matrix 
multiplication~\cite{LeGall2012}, and recall that $\omega\leq 3-\alpha$. 
In particular, from \eqref{eq:mm-exp} and $\Delta\geq 1$ we thus have
\begin{equation}
\label{eq:e1-cor2}
\begin{split}
1-\gamma+M(\Delta\gamma,\gamma)
&\leq 
1-\frac{\alpha}{2\Delta+\alpha}
+\frac{2\alpha\Delta}{2\Delta+\alpha}
+\frac{(\omega-2)\alpha}{2\Delta+\alpha}\\
&
=\frac{2\Delta+\alpha-\alpha+\alpha(2\Delta+\omega-2)}{2\Delta+\alpha}\\
&
=\frac{2\Delta(\alpha+1)+\alpha(\omega-2)}{2\Delta+\alpha}\\
&
\leq\frac{2\Delta(\alpha+1)+\alpha(1-\alpha)}{2\Delta+\alpha}\\
&
\leq\frac{4\Delta}{2\Delta+\alpha}\,.
\end{split}
\end{equation}
Furthermore, by definition of $\alpha$ and our choice of $\gamma$, we have
\begin{equation}
\label{eq:e2-cor2}
\begin{split}
M(1-\gamma,2\Delta\gamma)
&=M\biggl(\frac{2\Delta}{2\Delta+\alpha},\frac{2\Delta\alpha}{2\Delta+\alpha}\biggr)
=\frac{4\Delta}{2\Delta+\alpha}\,.
\end{split}
\end{equation}
Thus \eqref{eq:e2-cor2} dominates \eqref{eq:e1-cor2}. 
Simplifying \eqref{eq:e2-cor2}, we obtain
$\frac{4}{2+\alpha\bigl(1-\log_\tau\rho\bigr)}\,.$
This establishes the running time for approximate detection 
in Corollary~\ref{cor:alpha}.
The running time for listing is immediate by our choice of $\gamma$.
\end{proof}

\subsection{Proof of Corollary~\ref{cor:light-bulb}}

We recall Corollary~\ref{cor:light-bulb}:

\corlightbulb*

\begin{proof}
Fix a constant $0<\epsilon<\omega/3$ and let $n^{-\Theta(1)}<\rho<1$ be given. 
With the objective of eventually applying Theorem~\ref{thm:main}, 
let us begin by selecting a suitable value of $\tau$.
First, we want a small enough $0<\delta<1$ so that 
\[
\frac{2\omega}{3-\delta}\leq\frac{2\omega}{3}+\epsilon\,,
\]
or equivalently,
\begin{equation}
\label{eq:delta-upper-bound}
\delta \leq \frac{9\epsilon}{2\omega+3\epsilon}=\frac{9}{2\omega/\epsilon+3}\,.
\end{equation}
To obtain $\log_\tau\rho\leq\delta$, let us take $\tau=\rho^{1/\delta}$.
Because $0<\delta<1$ and $0<\rho<1$, we have $\tau<\rho$.
Let us now check that for all large enough $d$, all the inner products
between pairs of vectors in the input, except for the planted pair,
are at most $\tau d$ in absolute value with probability $1-o(1)$. For 
$d\geq 5\rho^{-2/\delta}\log n$
it follows from the Hoeffding bound
\eqref{eq:hoeffding} and the union bound that with probability $1-o(1)$ 
all of the at most $n^2$ pairs of vectors that have at least one
independent uniform random vector in the pair have inner product 
at most $\tau d$ in absolute value. Indeed, as $n\rightarrow\infty$ we have
\[
\begin{split}
2n^2\exp\biggl(-\frac{2(\tau d)^2}{4d}\biggr)
&=2n^2\exp\biggl(-\frac{\tau^2 d}{2}\biggr)
\leq 2n^2\exp\biggl(-\frac{5}{2}\log n\biggr) \rightarrow 0\,.
\end{split}
\]
It remains to find the planted pair with inner product 
at least $\rho d$ in absolute value. Let us apply the approximate detection 
algorithm in Theorem~\ref{thm:main}. To parameterize the algorithm, 
take $\Delta=1/(1-\delta)$ and $\gamma=1/(2\Delta+1)$. Observe in 
particular that both $\Delta$ and $\gamma$ depend only on the constants 
$\delta$ and $\epsilon$, and do not depend on $\rho$ or $\tau$.
Mimic the analysis in the proof of Corollary~\ref{cor:omega} 
to obtain from \eqref{eq:main-bound} the running time bound
\begin{equation}
\label{eq:tau-time}
\tilde O(n^{2\omega/3+\epsilon}\tau^{-4})\,.
\end{equation}
Observe furthermore that within the same time bound \eqref{eq:tau-time}
we can run the approximate detection algorithm recursively $O(\log\log n)$ times
on the pair of blocks marked by the algorithm to find the planted pair, w.h.p.
We obtain \eqref{eq:light-bulb-bound} by combining 
\eqref{eq:tau-time} with $\tau=\rho^{1/\delta}$ and 
\eqref{eq:delta-upper-bound}. 
\end{proof}

\subsection{Proof of Corollary~\ref{cor:sparse-parity}}
\label{sect:proof-sparse-parity}

We recall Corollary~\ref{cor:sparse-parity}:

\corsparseparity*

\begin{proof}
This proof relies on a split-and-list idea outlined by 
\cite[p.~32]{Valiant2015}; we present a proof here 
for completeness of exposition.

Let us start by setting up some notation. For a subset $A\subseteq[n]$
and a vector $(x_1,x_2,\ldots,x_n)\in\{-1,1\}^n$, let us write
$x_A=\prod_{\ell\in A}x_\ell$. (For the empty set, define $x_\emptyset=1$.)
Observe that for all $A,B\subseteq[n]$ it holds that
$x_Ax_B=x_{A\oplus B}$, where $A\oplus B=(A\setminus B)\cup (B\setminus A)$ 
is the symmetric difference of $A$ and $B$.
Let us write $\binom{[n]}{k}$ for the set of all $k$-subsets of~$[n]$.

Fix a constant $0<\epsilon<\omega/3$. Select a corresponding 
constant $0<\delta<1$ so that \eqref{eq:delta-upper-bound} holds.
Let $0<\eta<1$ be the given noise rate with $n^{-\Theta(1)}<|1-2\eta|<1$.
Let $S\subseteq[n]$ be the support of the parity function, $|S|=k\geq 2$.
Our task is to determine $S$ by drawing examples.
Let us draw $d$ examples $(x,y)\in\{-1,1\}^d\times\{-1,1\}$.
(We will fix a lower bound for $d$ in what follows.)
Recall that each label $y$ in an example has the structure $y=zx_S$, 
where $z\in\{-1,1\}$ independently with $\Pr(z=-1)=\eta$. 

Define a collection of 
$\binom{n}{\lfloor k/2\rfloor}+\binom{n}{\lceil k/2\rceil}\leq 2n^{(k+1)/2}$ 
vectors of dimension $d$ as follows. First, for each 
$J_1\in\binom{n}{\lfloor k/2\rfloor}$, construct the vector $a_{J_1}$ 
whose entries are the values $x_{J_1}$, where $x$ ranges over 
the $d$ examples $(x,y)$ that we have drawn. 
Second, for each \mbox{$J_2\in\binom{n}{\lceil k/2\rceil}$}, 
construct the vector $b_{J_2}$ whose entries are the values 
$x_{J_2}y$, where $x$ and $y$ range over the $d$ examples $(x,y)$
that we have drawn.

Let us write $\mathrm{Bin}_{\pm 1}(d,\beta)$ for the 
sum of $d$ random variables, each independently taking values 
in $\{-1,1\}$ such that $\beta$ is the probability of 
taking the value $-1$.

Let us study the inner products between the vectors in our collection.
We will use the notation ``$\sum_{(x,y)}$'' to indicate a sum
over the $d$ examples $(x,y)$ that we have drawn.
Observe that for all $J_1,J_1'\in\binom{n}{\lfloor k/2\rfloor}$ we have 
\begin{equation}
\label{eq:j1}
\bra a_{J_1},a_{J_1'}\ket=\sum_{(x,y)} x_{J_1}x_{J_1'}=\sum_{(x,y)} x_{J_1\oplus J_1'}\,. 
\end{equation}
Similarly, for all $J_2,J_2'\in\binom{n}{\lceil k/2\rceil}$
we have 
\begin{equation}
\label{eq:j2}
\bra b_{J_2},b_{J_2'}\ket=\sum_{(x,y)} x_{J_2}yx_{J_2'}y=\sum_{(x,y)} x_{J_2\oplus J_2'}\,. 
\end{equation}
Finally, for all $J_1\in\binom{n}{\lfloor k/2\rfloor}$
and $J_2\in\binom{n}{\lceil k/2\rceil}$ we have
\begin{equation}
\label{eq:j1j2}
\begin{split}
\bra a_{J_1},b_{J_2}\ket&=\sum_{(x,y)} x_{J_1}x_{J_2}y
=\sum_{(x,y)} x_{J_1}x_{J_2}x_Sz
=\sum_{(x,y)} x_{J_1\oplus J_2\oplus S}z\,. 
\end{split}
\end{equation}
Recalling that the vector $x$ in each example is an independent 
uniform random vector in $\{-1,1\}^n$, we observe%
\footnote{Indeed, for all {\em nonempty} $K\subseteq[n]$, the random variables
$\sum_{(x,y)}x_K$ and $\sum_{(x,y)}x_Kz$ have distribution 
$\mathrm{Bin}_{\pm 1}(d,1/2)$. For the symmetric difference it holds that 
$K_1\oplus K_2=\emptyset$ if and only if $K_1=K_2$.}
from \eqref{eq:j1}, \eqref{eq:j2}, and \eqref{eq:j1j2} that each
of the following inner products has distribution $\mathrm{Bin}_{\pm 1}(d,1/2)$:
(i) $\bra a_{J_1},a_{J_1'}\ket$ for distinct $J_1,J_1'$,
(ii) $\bra b_{J_2},b_{J_2'}\ket$ for distinct $J_2,J_2'$, and
(iii) $\bra a_{J_1},b_{J_2}\ket$ for all $J_1,J_2$ with $J_1\oplus J_2\neq S$.
Furthermore, all the {\em remaining} inner products between distinct vectors 
in our collection, that is, $\bra a_{J_1},b_{J_2}\ket$ for all $J_1,J_2$ 
with $J_1\oplus J_2=S$, have distribution $\mathrm{Bin}_{\pm 1}(d,\eta)$.
This difference between distributions enables us to detect a pair $J_1,J_2$ 
with $J_1\oplus J_2=S$ and hence the set $S$.

Take $\rho=|1-2\eta|$ and observe that $n^{-\Theta(1)}<\rho<1$. 
Let us take $\tau=\rho^{1/\delta}$. Because $0<\delta<1$ and $0<\rho<1$, 
we have $\tau<\rho$. Furthermore, $\log_\tau\rho\leq\delta$.
For 
\begin{equation}
\label{eq:d-lower-bound-2}
d\geq (2k+3)\rho^{-2/\delta}\log n 
\end{equation}
it follows from the Hoeffding bound
\eqref{eq:hoeffding} and the union bound that with probability $1-o(1)$ 
all of the at most $2n^{k+1}$ inner products with distribution 
$\mathrm{Bin}_{\pm 1}(d,1/2)$ are at most $\tau d$ in absolute value. 
Indeed, as $n\rightarrow\infty$ we have
\begin{equation}
\label{eq:binprob}
\begin{split}
4n^{k+1}\exp\biggl(-\frac{2(\tau d)^2}{4d}\biggr)
&=4n^{k+1}\exp\biggl(-\frac{\tau^2 d}{2}\biggr)\\
&\leq 4n^{k+1}\exp\biggl(-\frac{2k+3}{2}\log n\biggr) = 4n^{-1/2} \rightarrow 0\,.
\end{split}
\end{equation}

Observe that the expectation of $\mathrm{Bin}_{\pm 1}(d,\eta)$ is
$(1-2\eta)d$ and that the absolute value of the expectation is $\rho d$.
For $\eta\geq 1/2$ (respectively, for $\eta\leq 1/2$) 
the probability that $\mathrm{Bin}_{\pm 1}(d,\eta)$ is at most 
(respectively, at least) its expectation is at least $1/4$
\cite[Theorem~1 and Corollary~3]{Greenberg2014}.

We want to consider two events over the $d$ examples drawn: (a) that
the absolute value of at least one inner product with distribution
$\mathrm{Bin}_{\pm 1}(d,\eta)$ is at least $\rho d$, and (b) all of the
  at most $2n^{k+1}$ inner products with distribution
  $\mathrm{Bin}_{\pm 1}(d,1/2)$ are at most $\tau d$. By above, (a) occurs with at least probability 1/4, and, by \eqref{eq:binprob}, (b) occurs with probability $1-o(1)$.

Let us apply the approximate detection algorithm in Theorem~\ref{thm:main}. 
To parameterize the algorithm, 
take $\Delta=1/(1-\delta)$ and $\gamma=1/(2\Delta+1)$. Observe in 
particular that both $\Delta$ and $\gamma$ depend only on the constants 
$\delta$ and $\epsilon$, and do not depend on $\rho$ or $\tau$.
Mimic the analysis in the proof of Corollary~\ref{cor:omega} 
to obtain from \eqref{eq:main-bound} for all large enough $k$
the running time bound
\begin{equation}
\label{eq:tau-time-2}
\tilde O(n^{(\omega+\epsilon)k/3}\tau^{-4})\,.
\end{equation}
Observe furthermore that within the same time bound \eqref{eq:tau-time-2}
we can run the approximate detection algorithm recursively $O(\log\log n)$ 
times on any one pair of blocks marked by the algorithm to find at least
one inner product $\bra a_{J_1},b_{J_2}\ket$
with absolute value at least $\rho d$, and hence
the set $S=J_1\oplus J_2$ with probability 
$1-o(1)$. Here we have of course conditioned
on both events (a) and (b) above. 

The time bound also allows us to increase the odds of success to
$1-o(1)$ by performing $O(\log\log n)$ repetitions of the approximate
detection algorithm, each time with new $d$ examples. Indeed, since
the input is stochastic in nature, we have that (a) fails at most at 
probability $(3/4)^{\log\log n}$. Likewise, from \eqref{eq:binprob},
we have the bound $O(n^{-1/2})$ on the failure probability of
(b). Hence, by the union bound, the total probability of input that
causes the approximate detection algorithm to fail is bounded by
$(3/4)^{\log\log n}+\frac{\log\log n}{n^{-1/2}} \to 0$, so the
\emph{input} enables us to correctly detect the parity with
probability $1-o(1)$. It should also be noted that due to randomness
in the approximate detection algorithm itself, the \emph{algorithm}
also succeeds with probability $1-o(1)$.

We obtain \eqref{eq:sparse-parity-bound} by combining 
\eqref{eq:tau-time-2} with $\tau=\rho^{1/\delta}$ and 
\eqref{eq:delta-upper-bound}. Similarly, we obtain 
\eqref{eq:sparse-parity-examples} by combining 
\eqref{eq:d-lower-bound-2} with \eqref{eq:delta-upper-bound}.
\end{proof}

\subsection{Proof of Theorem~\ref{thm:main-two-level}}

We recall Theorem~\ref{thm:main-two-level}:

\thmmaintwolevel*

\begin{proof}
The proof is otherwise identical to the proof of Theorem~\ref{thm:main}
in \S\ref{sect:algorithm}, with the exception that we modify the listing 
phase as follows. For each of the at most $n^2$ pairs of blocks 
$k_1,k_2\in[\bar n/t]$ marked by the detection algorithm, we run the 
detection algorithm again, with parameter $\kappa$ replacing the parameter
$\gamma$, and using the same value for the parameter $\Delta$.
Indeed, each signalled pair of blocks can be viewed as an input of 
size $n^\gamma$ and dimension $d$. From these inputs we obtain as 
output at most $q$ pairs of blocks, each of size $n^{\gamma\kappa}$. 
Finally, we run the listing algorithm for these 
at most $q$ blocks. The running time and success probability follow 
immediately from Theorem~\ref{thm:main} and the observation that the 
number of recursive invocations on inputs of size $n^\gamma$ is 
at most $q\leq n^2$, so by the union bound the conclusion holds 
w.h.p.~in $n$. 
\end{proof}

\subsection{Proof of Corollary~\ref{cor:omega-two-level}}

We recall Corollary~\ref{cor:omega-two-level}:

\coromegatwolevel*

\begin{proof}
Analogous to Corollary~\ref{cor:omega}, take $\gamma=\kappa=1/(2\Delta+1)$.
\end{proof}

\subsection{Proof of Corollary~\ref{cor:alpha-two-level}}

We recall Corollary~\ref{cor:alpha-two-level}:

\coralphatwolevel*

\begin{proof}
Analogous to Corollary~\ref{cor:alpha}, take $\gamma=\kappa=\alpha/(2\Delta+\alpha)$.
\end{proof}

\section{A Lower Bound via Orthogonal Vectors}

\label{sect:orthogonal-vectors}

This section presents a local transformation from the orthogonal 
vectors problem (cf.~\cite{Abboud2015,Williams2014}) to 
\textsc{Bichromatic Out\-lier Cor\-re\-la\-tions}. We present this 
transformation for completeness of exposition only and remark that 
it has been superseded by recent results of Ahle, Pagh, Razenshteyn, and 
Silvestri~\cite[Theorem~2]{Ahle2015}. 
First, let us recall the orthogonal vectors problem:

\begin{Prob}[\textsc{Orthogonal Vectors}]
\ \ Given as input a set of $n$ vectors with dimension $d$
and entries in $\{0,1\}$, decide whether there is an orthogonal
pair of vectors over the integers.
\end{Prob}

The sparsification lemma of Impagliazzo, Paturi, and 
Zane~\cite{Impagliazzo2001b} and a lemma of Williams~\cite{Williams2005}
yield the following conditional hardness result
(cf.~\cite[Lemma~A.1]{Williams2014}) for \textsc{Orthogonal Vectors}:

\begin{lemma}[Williams~\cite{Williams2005}]
\label{lem:orthogonal-vectors}
Suppose there exists a constant $\delta>0$ and a randomized algorithm that
for all constants $c\geq 1$ solves the \textsc{Orthogonal Vectors} problem 
with $d\leq c\log n$ in time $O(n^{2-\delta})$,
w.h.p. Then, the Strong Exponential Time Hypothesis is false. 
\end{lemma}

The Orthogonal Vectors Conjecture (OVC) states that the algorithm assumed 
in Lemma~\ref{lem:orthogonal-vectors} does not exist 
(cf.~\cite{Abboud2015,Williams2014}).
Lemma~\ref{lem:orthogonal-vectors} together with the following local
transformation shows that \textsc{Bichromatic Out\-lier Cor\-re\-la\-tions} cannot be
solved in subquadratic time for inputs with $\log_\tau\rho$
arbitrarily close to $1$ unless OVC and subsequently SETH are false:

\begin{lemma}
\label{lem:local-trans}
Suppose that there exists a constant $\delta>0$ and a randomized
algorithm that for all constants $c\geq 1$ solves the 
\textsc{Bichromatic Out\-lier Cor\-re\-la\-tions} problem for Boolean
inputs with $d\leq c\log n$ and $|\rho-\tau|\leq 2/d$ in time 
$O(n^{2-\delta})$, w.h.p. Then, there exists a constant $\delta'>0$ 
and a randomized algorithm that for all constants $c'\geq 1$ solves the 
\textsc{Orthogonal Vectors} problem with $d'\leq c'\log n$ 
in time $O(n^{2-\delta'})$, w.h.p.
\end{lemma}

\begin{proof}
Without loss of generality we may work with a bichromatic version 
of \textsc{Ortogonal Vectors} where our input is a pair of matrices 
$S,T\in\{0,1\}^{d'\times n}$. We transform the matrices $S,T$ into a pair 
of matrices $A,B\in\{-1,1\}^{(4d'+1)\times n}$ such that for 
all $j_1,j_2\in[n]$ it holds that $|\bra a_{j_1},b_{j_2}\ket|$ is 
above a threshold value if and only if $\bra s_{j_1},t_{j_2}\ket=0$. 

The local transformation is given by a function $h :\{0,1\}^2 \to \mathbb Z$ 
that decomposes into two functions $u,v : \{0,1\}\to\{-1,1\}^k$ 
such that $h(x,y) = \bra u(x),v(y) \ket$ with $h(0,0)=h(1,0)=h(0,1)=k_1$
for some integer $k_1>0$, and $h(1,1) = k_2$ for some integer
$k_2<0$. 
One such pair of functions with $k=4$, $k_1 = 2$, and $k_2 = -2$ 
is the following:
\begin{align*}
u(x) & = \begin{bmatrix}1 & 1-2x & \hspace*{9mm}1 & 1-2x\end{bmatrix}^\top\,, \\
v(y) & = \begin{bmatrix}1 & \hspace*{9mm}1 & 1-2y & 2y-1\end{bmatrix}^\top\,.
\end{align*}
Apply the function $u$ (respectively, the function $v$) to each element 
of $S$ (respectively, $T$) to produce a matrix $A$ (respectively, $B$) 
of size $4d'\times n$. Finally, append one full row of 1-entries 
to both matrices $A$ and $B$.

The matrices $A$ and $B$ yield a $\{-1,1\}$-valued
input to \textsc{Out\-lier Cor\-re\-la\-tions} with the following parameters: 
the number of vectors is $n$, the dimension is $d=4d'+1$, the outlier parameter 
is $\rho=(2d'+1)/(4d'+1)$, and the background parameter $\tau=(2d'-1)/(4d'+1)$.
In particular, we observe that the inner products $\bra a_{j_1},b_{j_2}\ket$
lie in the interval $[-2d'+1,2d'+1]$, and we have 
$|\bra a_{j_1},b_{j_2}\ket|\geq \rho (4d'+1)$ if and only if 
$\bra s_{j_1},t_{j_2}\ket=0$.
Furthermore, without loss of generality we may assume that the
instance $A,B$ has at most $q=O(n^{2-\eta})$ outlier pairs
for a constant $\eta>0$. Indeed, we may test uniformly at
random $n^{1.1\eta}$ pairs of vectors for outlier pairs. 
If we find an outlier pair, we are done; otherwise we can conclude 
that w.h.p.~in $n$ the instance has at most $q$ outlier pairs.
In the latter case we run the assumed algorithm for 
\textsc{Out\-lier Cor\-re\-la\-tions} to find a pair if it exists. 
Observe that for our choice of $\rho,\tau$ we have 
$|\rho-\tau|=2/(4d'+1)=2/d$.
\end{proof}

\begin{remark}
We can replace the assumption $|\rho-\tau|\leq 2/d$ in 
Lemma~\ref{lem:local-trans} with
\[
\log_\tau\rho\geq
\frac{\log\frac{2d'+1}{4d'+1}}{\log\frac{2d'-1}{4d'+1}}
=1-\frac{\log\bigl(1+\frac{4}{d-3}\bigr)}{\log\bigl(2+\frac{6}{d-3}\bigr)}
=1-\Theta(1/d)\,.
\]
\end{remark}

\section*{Acknowledgement}
The research leading to these results has received funding from the 
European Research Council 
under the European Union's Seventh Framework 
Programme (FP/2007-2013) / ERC Grant Agreement 338077
``Theory and Practice of Advanced Search and Enumeration.''

A preliminary conference abstract of this paper appeared in Robert 
Krauthgamer (Ed.), Proceedings of the 27th ACM-SIAM Symposium on
Discrete Algorithms (SODA 2016, Arlington, VA, January 10--12, 2016),
SIAM, Philadelphia, PA, 2016, pp.~1288--1305.

We thank the anonymous referees for their helpful comments and
Rasmus Pagh and Thomas Dybdahl Ahle for useful discussions.
Work done in part while the second author was visiting the Simons Institute
for the Theory of Computing.


\bibliographystyle{acm}
\bibliography{refs}
\end{document}